\documentclass{elsarticle}
\usepackage{lineno,hyperref}
\modulolinenumbers[5]

\usepackage{amsmath,amsthm}
\usepackage{amssymb,latexsym}
\usepackage{graphicx}
\usepackage[labelsep=endash]{caption}
\usepackage{mathptmx}
\usepackage{subcaption}
\theoremstyle{definition}

\usepackage{amsmath, amssymb, url}
\usepackage{xcolor}

\begin{document}

\begin{frontmatter}
\title{Nonlinear convection stagnation point heat transfer and MHD fluid flow in porous medium towards a permeable shrinking sheet}
\author{Rakesh Kumar}
\author{Shilpa Sood}
\address{Department of Mathematics, Central University of Himachal Pradesh, Dharamshala, INDIA}
\ead{rakesh@cuhimachal.ac.in}
\begin{abstract}
This investigation deals with the analysis of stagnation point heat transfer and corresponding flow features of hydromagnetic viscous incompressible fluid over a vertical shrinking sheet. The considered sheet is assumed to be permeable and subject to addition of stagnation point to control the generated vorticity in the boundary layer. The sheet is placed on the right side of the fluid saturated porous medium which is having permeability of specified form. Nonlinear convection waves in the flow field are realized due to the envisaged nonlinear relation between density and temperature. The equations governing the nonlinear convection boundary layer flow are modeled and simplified using similarity transformations. The economized equations are solved for numerical solutions by employing the implicit finite difference scheme also known as Keller-box method. The influence of the associated parameters of the problem on velocity and temperature distributions, skin friction and rate of heat transfer are presented through graphs and tables, and qualitatively discussed. The  study reveals that interaction among magnetic field, porous medium permeability and nonlinear convection parameters substantially enhance the solution range and thus endorse their control to sustain the boundary layer flow. 
\end{abstract}
\end{frontmatter}
\section{Introduction}
The analysis of stagnation point heat transfer and flow of hydromagnetic fluids over permeable shrinking surfaces ingrained in porous medium finds profuse pertinence in engineering and manufacturing gimmick such as  purification of crude oil, glass manufacturing, paper production, polymer sheets, MHD electrical power generation, magnetic material processing, shrinking film etc. (Mahapatra et al. \cite{mahapatra2014}). The Stagnation point flow is known as a classical problem in fluid mechanics, as stagnation points exist virtually on all the solid objects submerged in the fluids or when flow infringes on a solid body. The stagnation point experiences maximum pressure, and heat and mass transfer. The extension of the pioneering work of Hiemenz \cite{hiemenz1911} to stagnation point flow over moving or stretching/shrinking surfaces has attracted the diligence of many investigators. In the above mentioned flows, the flows due to the shrinking of sheet is a new type of flow with a common example of rising shrinking balloon. These flows are differnt from those over stretching sheet problems in the sense that the fluid is attracted towards a slot whereas a far field suction would be induced towards a stretching sheet (Miklav{\v c}i{\v c} and Wang \cite{miklavcic2006}). The study of the stagnation flow problems towards a shrinking sheet began with the work of Wang \cite{wang2008}. He concluded that larger shrinking rate do not confirm solutions and multiple solutions may be obtained for the two dimensional case. He further remarked that the flow structure is complicated by the non-alignment of the stagnation flow and shrinking sheet. Goldstein \cite{goldstein1965} in his paper remarked that a shrinking flow is essentially a backward type flow. The physics behind this phenomenon is that the vorticity generated at the surface of the shrinking sheet does not remain confined with in the boundary layer which in turn induces a reverse flow. Later, Mahapatra and Nandy \cite{mahapatranandy} in their stability analysis analyzed that similarity solution will exist if a suitable suction is imposed or a stagnation point is added on the shrinking sheet. In our present analysis, we have also considered both suction and stagnation point to derive the invariant solutions. Ishak et al. \cite{ishak2010} has investigated the stagnation point flow over a shrinking sheet in a micropolar fluid and Fan et al. \cite{fan2010} analyzed the effect of unsteadiness on the stagnation point flow over a shrinking surface. Later on, Bhattacharyya et al. \cite{bhattacharyya2011} examined the slip effects on the stagnation point flow and the related heat transfer over a shrinking surface; and Rosca and Pop \cite{rosca2013} considered the second order slip for the stagnation flow over a vertical permeable shrinking surface. The similarity transformations in exponential form have been utilized by Bhattacharyya and Vajravelu \cite{bhattacharyya2012} to obtain the numerical solutions for the stagnation point flow and heat transfer over an exponentially shrinking sheet and Bachok et al. \cite{bachok2012} extended it for nanofluids. Some contributions on the study of stagnation-point flow over shrinking surfaces for various physical situations and different fluids appeared in literature. Some of them can be found in Yian et al. \cite{yian2011}, Bachok et al. \cite{bachok2013}, Nadeem et al. \cite{25}, Bhattacharyya \cite{26}, Gorder et al. \cite{27},  Mansur et al. \cite{28} and Ishak et al. \cite{29}. \\

In the above mentioned papers, either free/forced convection or mixed convection effects were considered to discuss the flow  and heat transfer characteristics. But, when the temperature difference between the surface and ambient fluid is substantially large, the nonlinear density temperature (NDT) variations in the buoyancy force term cannot be ignored as the distribution of flow and heat transfer is significantly affected by it (Vajravelu and Sastri \cite{vajravelu1976}). In their study, they suggested that we can consider the variation of density with respect to temperature as 
$$
\rho(T)=\rho(T_s)+\left(\frac{\partial \rho}{\partial T}\right)_s\left(T-T_s\right)+\left(\frac{\partial^2 \rho}{\partial T^2}\right)_s\left(T-T_s\right)^2+...\nonumber
$$
where $T_s$ speaks for surface temperature.\\
If we retain the terms upto second order, the above equation can be written as
$$
\Delta \rho/\rho=-\beta_0 \left(T-T_s\right)-\beta_1 \left(T-T_s\right)^2
$$
which may be called as the NDT variation.\\
The above relation also accommodates the linear temperature dependence and is also apropos to the analysis of the flow of water at $4^o$C. Mekker et. al. \cite{mekker1979} analyzed maximum density effects on the onset of convection in a horizontal layer of water. Due to the significance of NDT in larger temperature differences, we shall be having a parameter $\gamma=\frac{\beta_1 \Delta T}{\beta_0}$ in our analysis and would be called as nonlinear convection parameter. In this sequence, one paper of Bhargava and Agarwal \cite{bhargava1979} for the study of fully developed free convection flow with nonlinear density temperature variations in a circular pipe appeared in literature. Later on, the effect of nonlinear convection on the flow over a flat permeable plate was investigated by Vajravelu et al. \cite{vajravelu2003}.\\

Inspired by the significance of nonlinear convection in larger temperature differences and work of Vajravelu et al. \cite{vajravelu2003}, we have set the target to achieve the numerical solutions of the boundary layer stagnation point flow and heat transfer over a permeable shrinking sheet lying on the right side of the porous medium of uniform porosities. To the best of our knowledge, no other authors have considered the influence of nonlinear convection on the flow and heat transfer  characteristics interconnected with the stagnation point over a vertical permeable sheet ingrained in porous medium. Hence, our results are new and genuine. 
 
\section{Analysis of the problem}
Consider the steady two-dimensional flow of an electrically conducting viscous incompressible fluid near the stagnation point over a vertical permeable sheet which is shrinking in its own plane. The sheet is assumed to be on the right side of the porous medium of uniform porosities. The extended Darcy model has been exploited to represent porous medium. Here, $x$-axis is measured along the sheet and $y$ axis is taken normal to it. The origin of coordinate system and stagnation point is chosen as a common fixed point that remains unaltered by the shrinking forces which are acting from opposite directions towards the origin. A magnetic field of uniform strength $ B_0$ is applied normal to the palne of considered shrinking surface. The magnetic Reynolds number is reckoned to be small to ignore the induced magnetic field. We have also presumed that the external flow obtrudes on the shrinking surface with the velocity $U_e (x)=ax$, where $ a>0$ is the stagnation flow tenacity, and the flat surface will constrict with the velocity velocity $U_w (x)=cx$, where $ c<0$  represents shrinking of the sheet and $c>0$ represents stretching. The temperature of the sheet is taken as
$T_w \left(x\right)=T_\infty+bx$, where $b>0$ for $T_w \left(x\right)>T_\infty$ and $b<0$ for $T_w \left(x\right)<T_\infty$. \\
The governing boundary layer equations together with the above stated assumptions which govern the considered flow can be written as:\\
\begin{align}\label{1.3.1}
\frac{\partial u}{\partial x}+\frac{\partial v}{\partial y}=0
\end{align} 
\begin{align}\label{1.3.2}
 u\frac{\partial u}{\partial x}+v\frac{\partial u}{\partial y}=U_e  \frac{\partial U_e}{\partial x} +\nu  \frac{\partial^{2} u}{\partial y^{2} }+\frac{\nu}{k_0}\left(U_e-u\right)&+ \frac{\sigma B_0^{2}}{\rho }\left(U_e-u\right)& \nonumber \\ 
&+g\beta \left(T-T_\infty\right)+g\beta_1  \left(T-T_\infty\right)^{2}&
\end{align} 
\begin{align}\label{1.3.3}
u\frac{\partial T}{\partial x}+v\frac{ \partial T}{\partial y}=\alpha\frac{\partial^{2} T}{\partial y^{2 }}                                                                                                                    
\end{align} 
The boundary conditions relevant to the problem are:
\begin{align}
v=v_w(x),  \quad u=U_w\left (x\right)=cx, \quad T=T_w \left(x\right)=T_\infty+bx \qquad    &\mbox{at}& \quad y=0 \label{1.3.4}\\
u=U_e\left (x\right)=ax,   \quad    T=T_\infty       \qquad \qquad \qquad \qquad \qquad &\mbox{as}& y\rightarrow\infty \label{1.3.5}                                              
\end{align}
The momentum equation \eqref{1.3.2} and equation of energy \eqref{1.3.3} in their self-similar form are reduced to the following nonlinear coupled ordinary differential equations:
\begin{align}\label{1.3.7}
f'''\left(\eta\right)+f\left(\eta\right)f''\left( \eta\right)-\left[f'\left(\eta\right)\right]^{2}&+1+K\left[1  -f'\left(\eta\right)\right]&\nonumber\\
&+M\left[1  -f'\left(\eta\right)\right]+\lambda\theta\left(\eta\right)\left[1+\gamma\theta\left(\eta\right)\right] = 0&                                                         
\end{align}
\begin{align} \label{1.3.8}                     
\theta''\left(\eta\right) +Pr\left[f\left(\eta\right)\theta'\left(\eta\right)-\theta\left(\eta\right)f'\left(\eta\right)\right ]=0   
\end{align} 
The corresponding reduced boundary conditions are:
\begin{align} \label{1.3.9}
f\left(\eta\right) =S, \qquad     f'\left(\eta\right) = \frac{c}{a}=\epsilon, \qquad      \theta\left(\eta\right) =1   \qquad   \mbox{at}\qquad  \eta =0\\    
f'\left(\eta\right) =1, \qquad \theta\left(\eta\right)= 0                     \qquad \mbox{as}\qquad          \eta\rightarrow\infty  \label{1.3.10}
\end{align}
where $S=-\frac{v_w}{\sqrt{\nu a}}$, $S>0$ represents the suction of the fluid, and $\epsilon$ is the ratio of shrinking velocity rate to the straining velocity rate.\\                                                                                                                                                                                                 
In the above equations [\eqref{1.3.7} to \eqref{1.3.10}], the following similarity transformations have been exploited: 
\begin{align}\label{1.3.6}
\eta=\sqrt{\frac{a}{\nu}} y,\hspace{.3cm} \psi=\sqrt{av} xf(\eta) ,\hspace{.3cm} \theta\left(\eta\right)= \frac{T-T_\infty}{T_w-T_\infty}, \hspace{.3cm} u=\frac{\partial\psi}{\partial y},\hspace{.3cm}v=-\frac{\partial \psi}{\partial x  }   
\end{align}
During this process, following non-dimensional parameters are eventuated:
$K=\frac{\nu x}{k_0 U_e}$ (Permeability parameter), $ \quad M=\frac{\sigma B_0^{2}} {\rho a}$  ( magnetic parameter),\quad $Pr=\frac{\nu}{\alpha} (\mbox{Prandtl number})$,\quad $\lambda=\frac{Gr_x}{Re_x^{2}}$(buoyancy parameter),\quad $Gr_x=\frac{g\beta(T_w-T_\infty)x^{3}}{\nu^{2}}  (\mbox{local Grashoff number})$,\quad  $Re_x =\frac {xU_e}{\nu}  (\mbox{ local Reynolds number})$\quad and \quad $\gamma=\frac{\beta_1 (T_w-T_\infty)}{\beta}  ( \mbox{nonlinear convection parameter})$.\\
The quantities of engineering interest which are associated with this study are the skin friction coefficient $C_f $ and local Nusselt number $Nu_x$, and are defined as follows:
\begin{align} \label{1.3.11}
C_f= \frac{\tau_w}{\rho U_e^{2}}      \hspace{2cm} \mbox{and}                       \hspace{2cm} Nu_x=\frac{ xq_w}{k(T_w-T_\infty)}    
\end{align}   
where $\tau_w=\mu(\frac{\partial u}{\partial y})_{y=0}  $ and $ q_w=-k(\frac {\partial T}{\partial y})_{y=0}$ are shear stress and heat flux at the sheet respectively.\\
The converted skin friction coefficient and Nusselt number in terms of dimensionless velocity $f''(0)$ and dimensionless temperature $-\theta'(0)$ can be respectively obtained as: 
\begin{align}\label{1.3.12}
C_f Re_x^{1/2}=f''(0) \hspace{2cm} \mbox{and} \hspace{2cm} {Nu_x }{Re_x^{1/2}}=-\theta'(0)                                                                                                             
\end{align}
\section{Solution methodology}
The coupled nonlinear boundary value problem constituted by equations \eqref{1.3.7}-\eqref{1.3.8} and boundary conditions \eqref {1.3.9} and \eqref{1.3.10} are solved numerically by employing Keller-box method (\cite{33}) as follows: \\
Substituting $f'=u$, $u'=v$ and $\theta'=q$ in equations \eqref{1.3.7}-\eqref{1.3.8}  and writing the obtained system of first order equations in finite difference form by utilizing central difference for derivatives and average at mid-point for variables, we obtain the following set of difference equations:
\begin{align}
f_j-f_{j-1} -\frac{h_j}{2}\left(u_j+u_{j-1} \right)=0 \label{1.3.99} 
\end{align}
\begin{align}
u_j-u_{j-1}-\frac{h_j}{2} \left(v_j+v_{j-1} \right)=0
\end{align}
\begin{align}
\theta_j-\theta_{j-1}-\frac{h_j}{2} \left(q_j+q_{j-1} \right)=0
\end{align}
\begin{align}
&v_j-v_{j-1}+\frac{h_j}{4} \left(f_j+f_{j-1} \right)\left(v_j+v_{j-1} \right)-\frac{h_j}{4} \left(u_j+u_{j-1}\right )^2+h_j+ M h_j \left(1-\frac{\left(u_j+u_{j-1} \right)}{2}\right)& \nonumber\\
&+ K h_j \left(1-\frac{\left(u_j+u_{j-1} \right)}{2}\right)+\lambda\frac{h_j}{2} \left(\theta_j+\theta_{j-1} \right)\left(1+\gamma\frac{\left(\theta_j+\theta_{j-1} \right)}{2}\right)=0&
\end{align}
\begin{align}
q_j-q_{j-1}+Pr\frac{h_j}{4}\left[\left(f_j+f_{j-1}\right)\left( q_j+q_{j-1}\right )-\left(u_j+u_{j-1} \right)\left(\theta_j+\theta_{j-1} \right)\right]=0 \label{1.3.100} 
\end{align}
In the above system of equations $j=1,2,3...J-1$. The boundary conditions in finite difference form can be grouped as
\begin{align}
f_0=S,\qquad u_0=\epsilon\qquad \theta_0=1, \hspace{2cm} \mbox{and}\hspace{2cm} u_J=1, \qquad \theta_J=0
\end{align}
The system of equations \eqref{1.3.99} to \eqref{1.3.100} are coupled first order nonlinear difference equations and are linearized using Newton's linearization  scheme, that is, we assume
\begin{align}
f_j^{i+1}&=f_j^i+\delta f_j^i , \qquad u_j^{i+1}=u_j^i+\delta u_j^i , \qquad v_j^{i+1}=v_j^i+\delta v_j^i ,&\nonumber \\ \theta_j^{i+1}&=\theta_j^i+\delta \theta_j^i , \qquad q_j^{i+1}=q_j^i+\delta q_j^i & 
\end{align}
Now the obtained linearized set of equations can be written in matrix vector form as
\begin{align}\label{3.1.1A}
A\delta=r
\end{align}
where A will be a tridiagonal matrix whose entries are blocks of order $5\times5$.\\
The solution of equation \eqref{3.1.1A} can be obtained using block-elimination method, which consist of forward and backward sweeps. The detailed description of the method has been avoided in order to preserve the space.

\section{Results and discussion}
In order to understand the influence of pertinent physical parameters of the problem on the distributions of velocity and temperature profiles, the numerical results are obtained through comprehensive and rigorous
\begin{table}[!htbp] 
\caption{Comparison of the values of $f''(0)$ with $\gamma=\lambda=M=K=S=0$ for $\epsilon<0$  }\label{1.3.T1}
\begin{tabular}{c c c c c c c}
\hline
\multicolumn{1}{c}{} &
\multicolumn{2}{c}{Wang \cite{wang2008}}&
\multicolumn{2}{c}{Mahapatra et al.\cite{mahapatra2014}}&
\multicolumn{2}{c}{Present Result}\\
\cline{2-3}
\cline{4-5} 
\cline{6-7}

$\epsilon$&$1^{st}$ solution&$2^{nd}$ solution &$1^{st}$ solution &$2^{nd}$ solution &$1^{st}$ solution &$2^{nd}$ solution\\
\hline
$-0.25$&1.4022&--&1.4022&--&1.4022&--\\
$-0.50$&1.4956&--&1.4956&--&1.4956&--\\
$-0.75$&1.4893&--&1.4892&--&1.4893&--\\
$-1.00$&1.3288&0.0&1.3288&0.0&1.3288&0.0\\
$-1.10$&--&--&1.1866&0.0492&1.1867&0.0493\\
$-1.15$&1.0822&0.1167&1.0822&0.1167&1.0822&0.1172\\
$-1.20$&--&--&0.9324&0.2336&0.9325&0.2336\\
$-1.2465$&0.5843&--&0.5843&0.5542&0.5849&0.5537\\
\hline
\end{tabular}
\end{table} computations procedure involving symbolic language in MATLAB. The relative tolerance error for convergence of results has been decided as $10^{-5}$, and iterative procedure is continued until the difference between the present and previous iterations does not become less than $10^{-5}$. The results so obtained are presented through graphs and tables which are subsequently discussed with qualitative approach. The accuracy of the numerical results has been validated through the comparison from earlier reported work in the literature and is being displayed through Table \ref{1.3.T1}, which supports and affirms the
 accuracy of applied numerical method. Motivated by the precision of numerical results, the present analysis is carried out to understand the influence of governing physical parameters involved in this study of stagnation point flow due to shrinking of the permeable sheet. In the entire analysis the related parameters are assigned arbitrary values and we have chosen $M=0.2, K=1, S=0.01, \epsilon=-3, \lambda=0.15, \gamma=0.2, Pr=3$ as arbitrary values throughout. \\

\begin{table}[!htbp]
\caption{Bifurcation points of $f''(0)$ and $-\theta'(0)$ and dual solution range  for $M, K, S$ with $\gamma=\lambda=0$ }\label{1.3.T2}
\begin{tabular}{c c c c c c }
\hline
\multicolumn{1}{c}{}&
\multicolumn{1}{c}{} &
\multicolumn{1}{c}{} &
\multicolumn{2}{c}{Bifurcation point}&
\multicolumn{1}{c}{Dual solution range}\\
\cline{4-5}
$M$ & $K$ & $S$ & ($\epsilon_c$, $f''(0)$) & ($\epsilon_c$, $-\theta'(0)$) &($\epsilon_l,\epsilon_r$) \\
\hline
$0.2$&--&--&(-2.43, 1.170)&(-2.43, -60.300 )&[-2.43,-1.470]\\
$1.0$&--&--&(-3.19, 1.430)&(-3.19, 438.500)&[-3.19, -1.736]\\
$1.5$&--&--&(-3.67, 1.550)&(-3.67, 179.590)&[-3.67,-1.882]\\
$2.0$&--&--&(-4.16, 1.682)&(-4.16, 124.960)&[-4.16,-2.018]\\
--&$1.5$&--&(-3.67, 1.550)&(-3.67, 179.590)&[-3.67,-1.882]\\
--&$2.0$&--&(-4.16, 1.682)&(-4.16, 124.96)&[-4.16,-2.018]\\
--&$2.5$&--&(-4.64, 1.810)&(-4.64, 106.980)&[-4.64,-2.144]\\
--&$3.0$&--&(-5.13,1.930)&(-5.13, 99.170)&[-5.13,-2.265]\\
--&--&$0.1$&(-2.96, 1.220)&(-2.96, -35.740)&[-2.96,-1.633]\\
--&--&$0.3$&(-3.05, 1.540)&(-3.05, 47.760)&[-3.05,-1.715]\\
--&--&$0.4$&(-3.10, 1.690)&(-3.10, 22.750)&[-3.10,-1.757]\\
--&--&$0.5$&(-3.15, 1.860)&(-3.15, 15.270)&[-3.15,-1.799]\\
\hline
\end{tabular}
\end{table}
\par Tables \ref{1.3.T2} and \ref{1.3.T3} have been framed to sort out the dual solution range, and bifurcation points where first and second solutions
\begin{table}[!htbp]
\caption{Bifurcation points of $f''(0)$ and $-\theta'(0)$ and dual solution range  for $M, K, S$ with $\gamma\neq\lambda\neq0$}\label{1.3.T3}
\begin{tabular}{c c c c c c c }
\hline
\multicolumn{1}{c}{}&
\multicolumn{1}{c}{} &
\multicolumn{1}{c}{} &
\multicolumn{3}{c}{Bifurcation point}&
\multicolumn{1}{c}{Dual solution range}\\
\cline{4-6}
$M$ & $K$ & $S$ & $(\epsilon_c$, $f''(0))$ & $(\epsilon_c$, $-\theta'(0))$ &  $(\epsilon_c$, $-\theta'(0)$)&($\epsilon_l,\epsilon_r)$ \\
\hline
$0.2$&--&--&(-5.52, 5.997)&(-5.52, -9.249)&(-0.33, 1.43)&[-5.520, -0.280]\\
$1.0$&--&--&(-3.15, 6.618)&(-3.15, -2.791)&(-0.39, 1.435)&[-3.15, -0.302]\\
$1.5$&--&--&(-2.99, 6.981)&(-2.99, -2.448)&(-0.43, 1.421)&[-2.99, -0.327]\\
$2.0$&--&--&(-2.96, 7.472)&(-2.96, -1.972)&(-0.47, 1.400)&[-2.96, -0.360]\\
--&1.5&--&(-2.99, 6.981)&(-2.99, -2.448)&(-0.43, 1.421)&[-2.99, -0.327]\\
--&2.0&--&(-2.96, 7.472)&(-2.96, -1.972)&(-0.47, 1.4000)&[-2.96, -0.360]\\
--&2.5&--&(-2.98, 7.969)&(-2.98, -1.855 )&(-0.51, 1.38)&[-2.98, -0.390]\\
--&3.0&--&(-3.04, 8.419)&(-3.04, -2.080)&(-0.56, 1.363)&[-3.04, -0.429]\\
--&--&0.1&(-1.69, 4.986)&(-1.69, -0.388)&(-0.187, 1.448)&[-1.69, -0.131]\\
--&--&0.2&(-3.29, 6.500)&(-3.29, -3.1912)&(-0.375, 1.451)&[-3.29, -0.300]\\
--&--&0.3&(--, --)&(--, --)&(-0.563, 1.455)&(--, -0.454]\\
--&--&0.4&(--,--)&(--, --)&(-0.75, 1.463)&(--,-0.609]\\
--&--&0.5&(--, --)&(--, --)&(-0.935, 1.477)&(--, -0.760]\\
\hline
\end{tabular}
\end{table}
 meet. Since the significance of this analysis lies in understanding the influence of nonlinear convection parameter on the flow and heat transfer characteristics, so a comparative analysis has been presented to predict its effect on skin friction $f''(0)$ and Nusselt number $-\theta'(0)$. The comparative analysis consist of two cases where in first case free convection (i.e. $\lambda=\gamma=0$) has been reckoned and in the second case nonlinear convection has been considered (i.e. $\lambda\neq0, \gamma\neq0$). Here, we are pointing towards a study made by  Miklav{\v c}i{\v c} and Wang \cite{miklavcic2006} in which the
\begin{figure}[!htbp]
\centering
   \begin{subfigure}{1.2\linewidth} \centering
     \includegraphics[scale=.3]{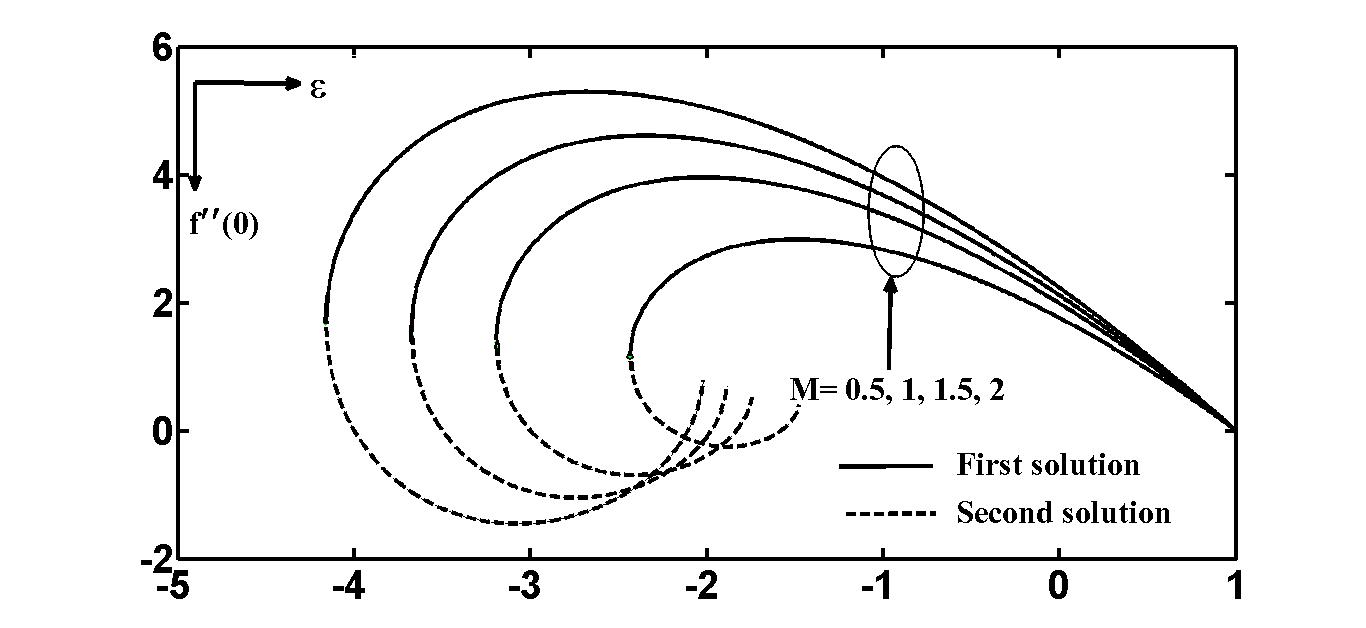}
     \caption{Skin friction for $\lambda=\gamma=0$} \label{1.3.2a}
   \end{subfigure}\\
   \begin{subfigure}{1.2\linewidth} \centering
     \includegraphics[scale=0.3]{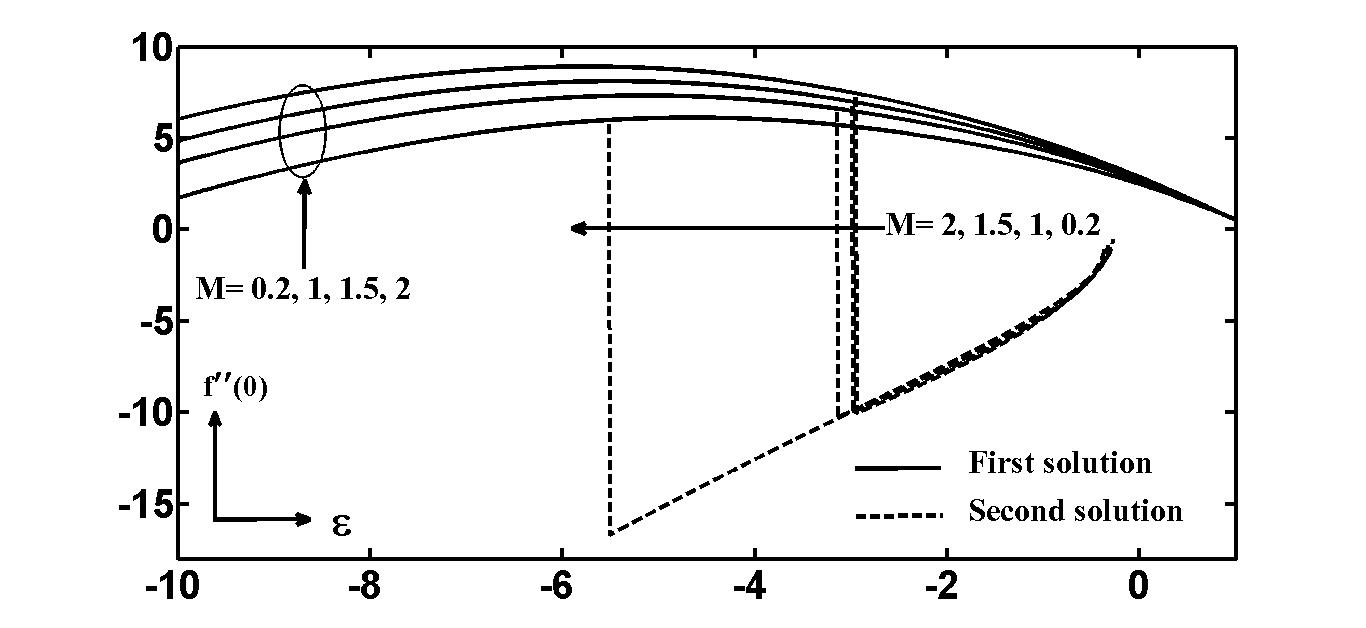}
     \caption{ Skin friction for $\lambda=2,\gamma=0.5$}\label{1.3.2b}
   \end{subfigure}
\caption{Variation of Skin friction with $\epsilon$ for various values of $M$ with $ S=0.2, K=1, Pr=3$ } \label{fig:1.3.2}
\end{figure} results reveal that multiple solutions exist for flows over shrinking surfaces and solution exist only for small shrinking rates. We have also observed dual solutions in our analysis as Table \ref{1.3.T2} and \ref{1.3.T3} indicate. The bifurcation points and dual solution range detected in Tables \ref{1.3.T2} and \ref{1.3.T3} evidently shows that dual solution range is elevated with the enhancement of magnetic field and porous medium permeability for the case when velocity field is independent from temperature field, whereas if we consider nonlinear convection waves in the flow field, the dual solution range is curtailed as Table \ref{1.3.T3} is indicating. Wang \cite{wang2008} claimed that dual solution exist for $-1.2465\leq \epsilon\leq -1$. Here interesting result which we observe from Table \ref{1.3.T2} is the significant increase in the dual solution range if effects of suction are included.\\

\begin{figure}[!htbp]
\centering
   \begin{subfigure}{1.2\linewidth} \centering
     \includegraphics[scale=0.4]{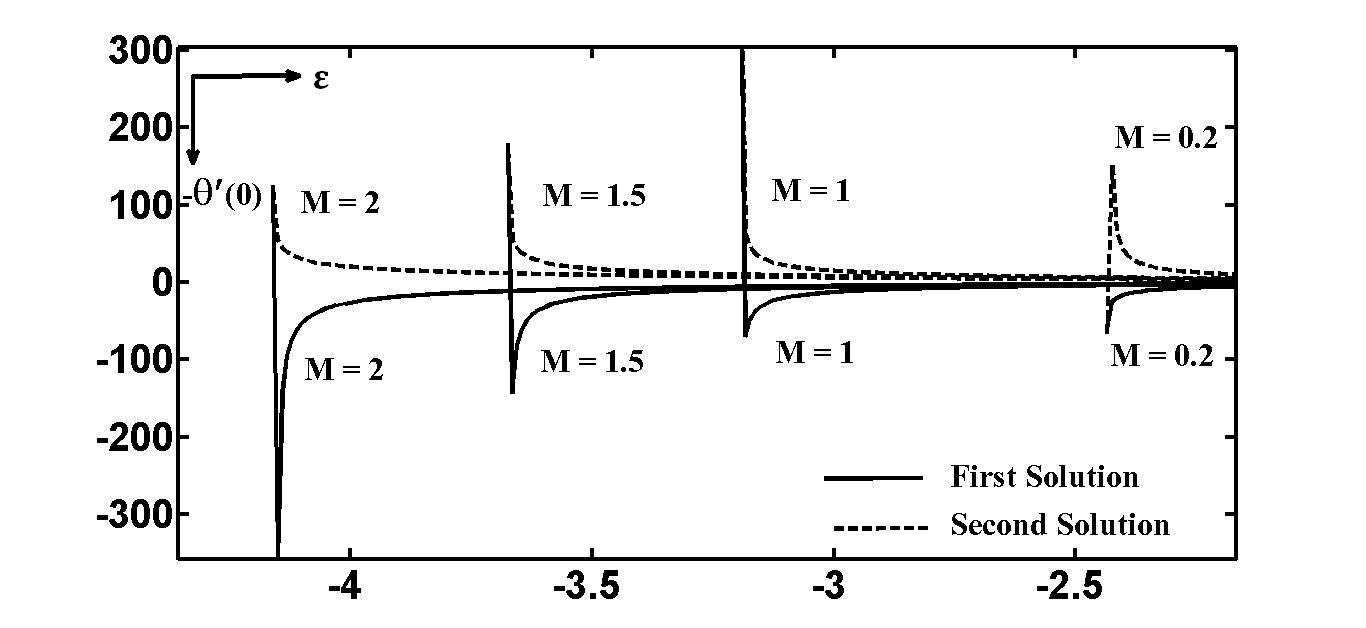}
     \caption{Nusselt number for $\lambda=\gamma=0$}\label{1.3.3a}
   \end{subfigure}\\
   \begin{subfigure}{1.2\linewidth} \centering
     \includegraphics[scale=0.3]{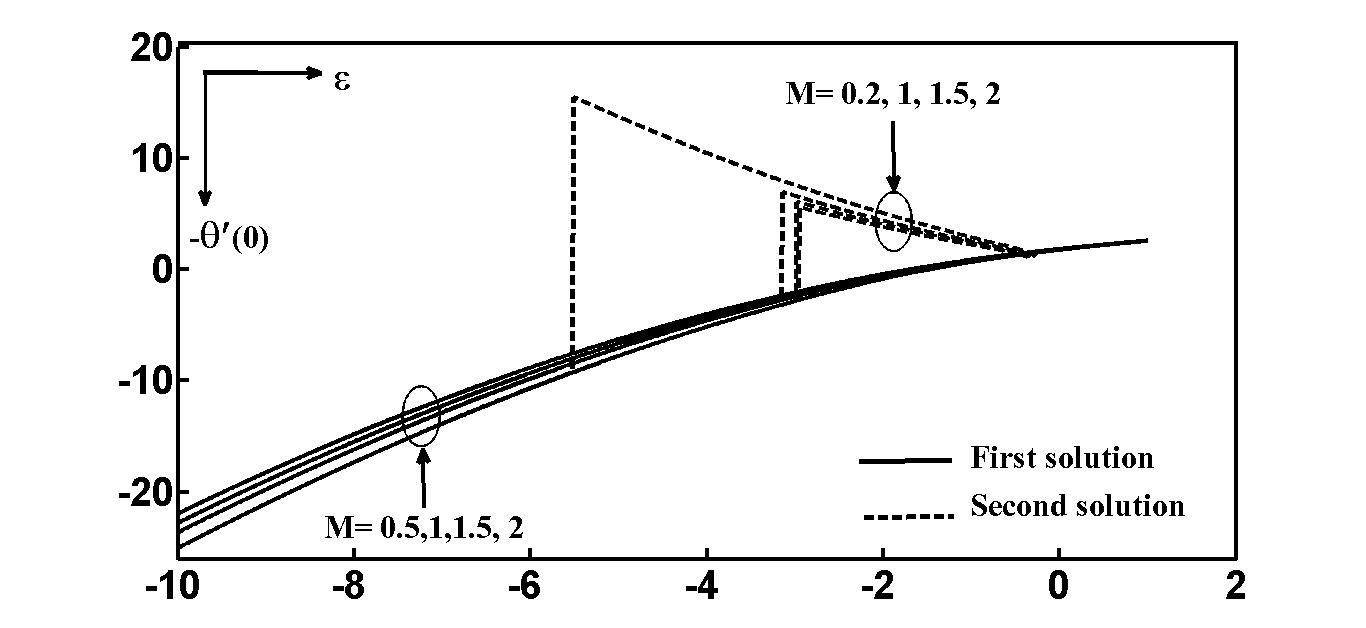}
     \caption{Nusselt number for $\lambda=2,\gamma=0.5$}\label{1.3.3b}
   \end{subfigure}
\caption{Variation of $-\theta'(0)$ with $\epsilon$ for various values of $M$ with $ S=0.2, K=1,Pr=3$}  \label{fig:1.3.3}
\end{figure}

\begin{figure}[!htbp]
\centering
   \begin{subfigure}{1.2\linewidth} \centering
     \includegraphics[scale=0.3]{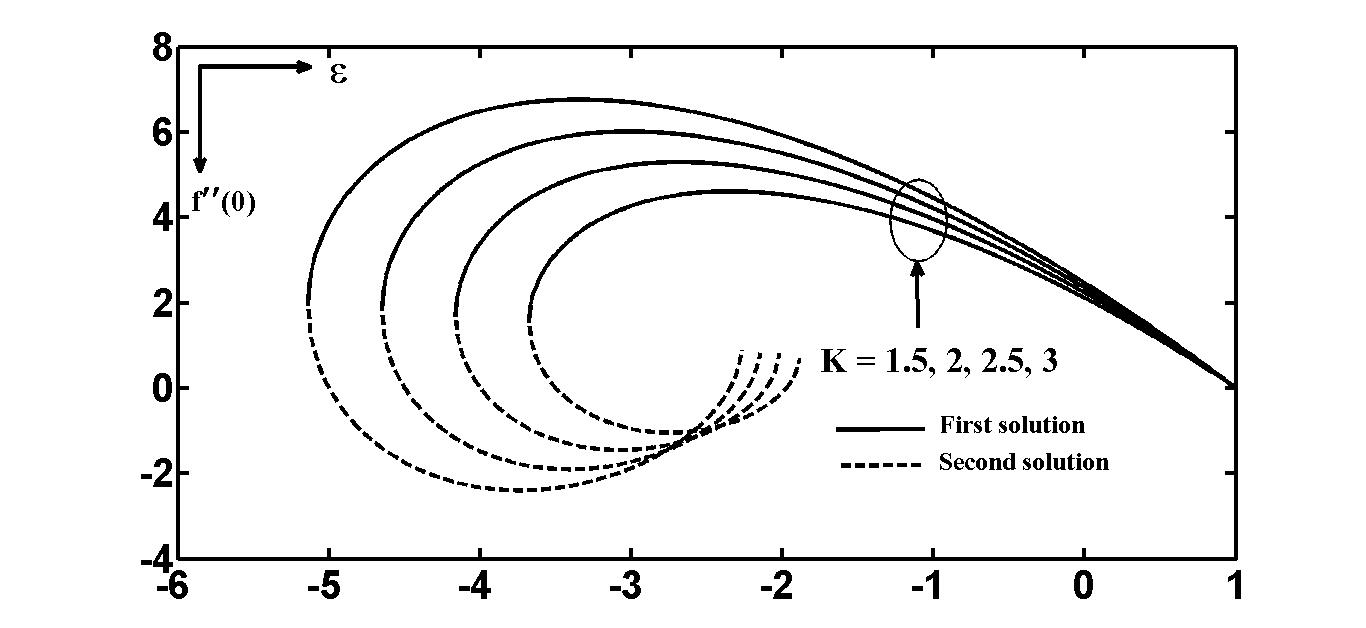}
     \caption{Skin friction for $\lambda=\gamma=0$} \label{1.3.4a}
   \end{subfigure}\\
   \begin{subfigure}{1.2\linewidth} \centering
     \includegraphics[scale=0.3]{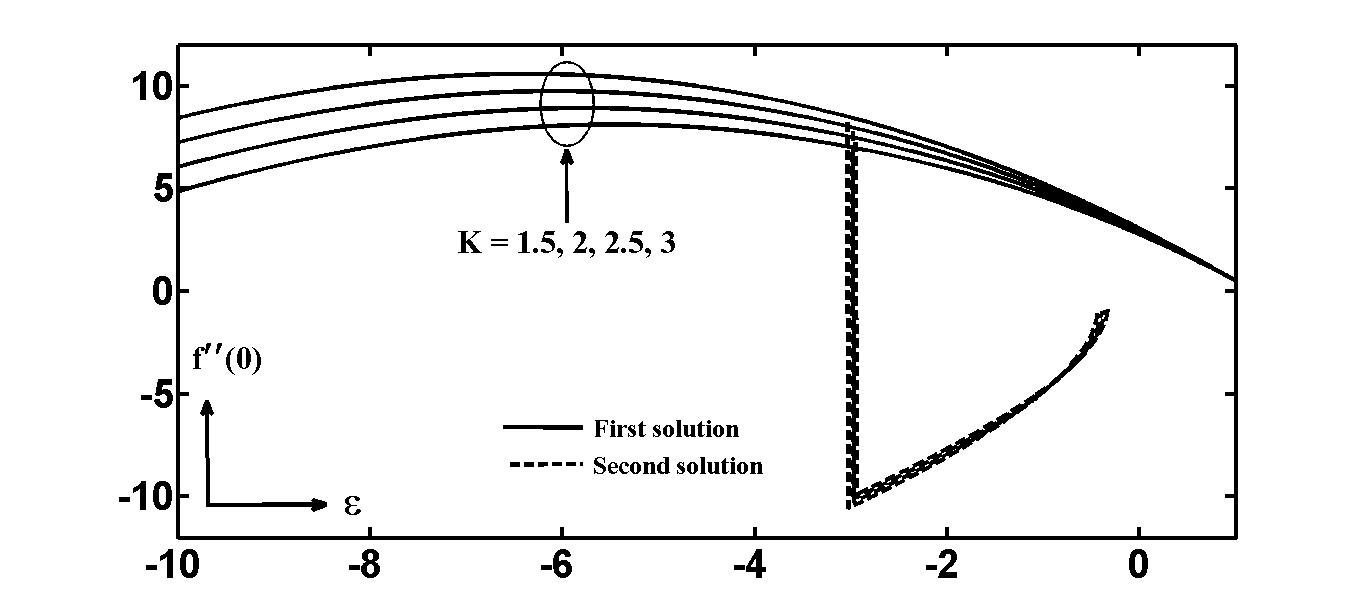}
     \caption{Skin friction for $\lambda=2,\gamma=0.5$}\label{1.3.4b}
   \end{subfigure}
\caption{ Variation of Skin friction with $\epsilon$ for various values of $K$ with $ S=0.2, M=1, Pr=3$}  \label{fig:1.3.4}
\end{figure}

\begin{figure}[!htbp]
\centering
   \begin{subfigure}{1.2\linewidth} \centering
     \includegraphics[scale=0.3]{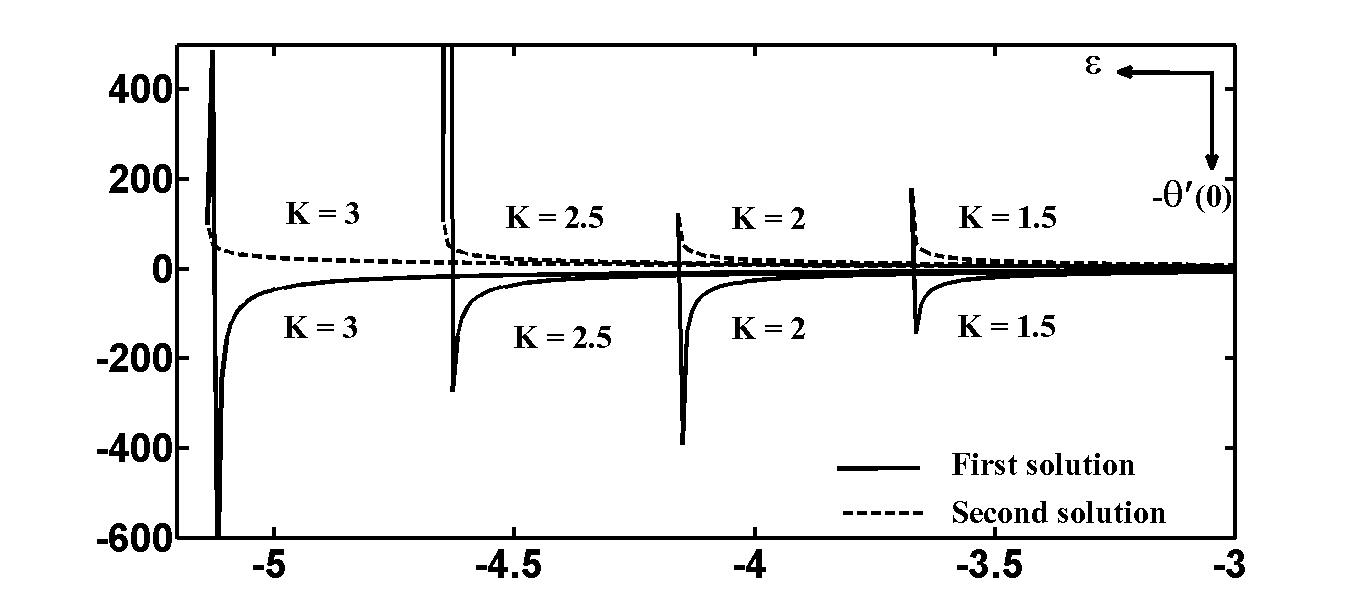}
     \caption{Nusselt number for $\lambda=\gamma=0$} \label{1.3.5a}
   \end{subfigure}\\
   \begin{subfigure}{1.2\linewidth} \centering
     \includegraphics[scale=0.3]{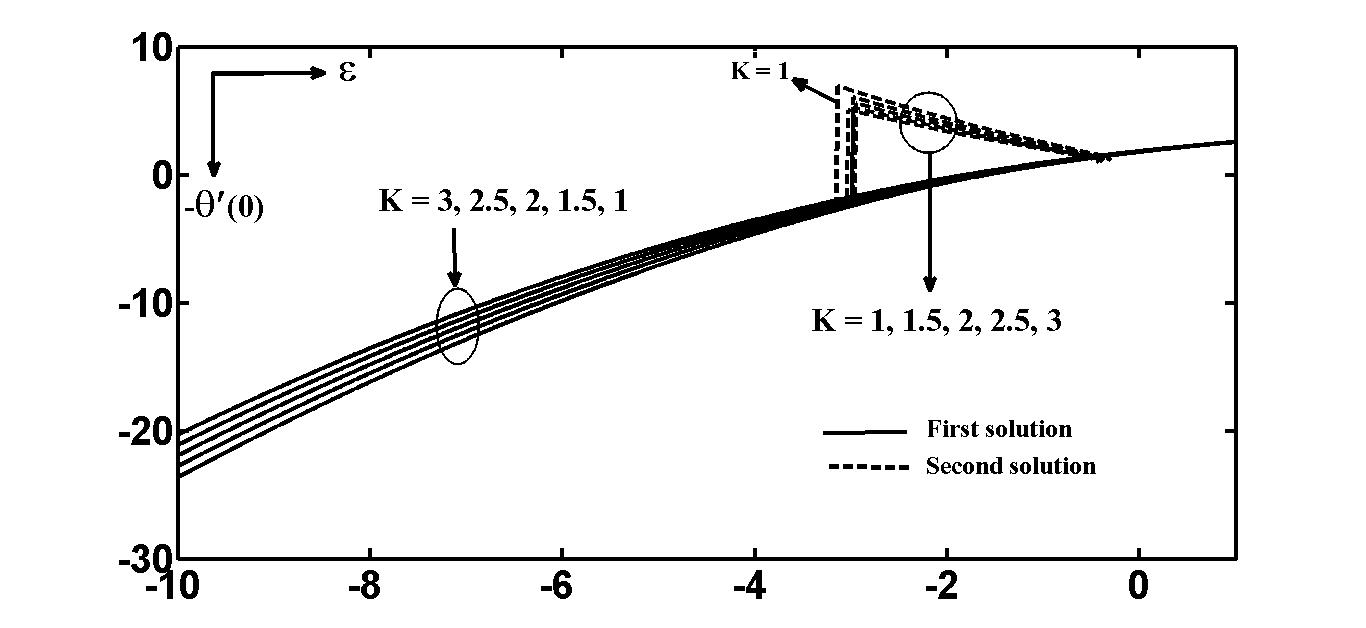}
     \caption{Nusselt number for $\lambda=2,\gamma=0.5$}\label{1.3.5b}
   \end{subfigure}
\caption{Variation of $-\theta'(0)$ with $\epsilon$ for various values of $K$ with $ S=0.2, M=1,Pr=3$} \label{fig:1.3.5}
\end{figure}
 
\begin{figure}[!htbp]
\centering
   \begin{subfigure}{1.2\linewidth} \centering
     \includegraphics[scale=0.3]{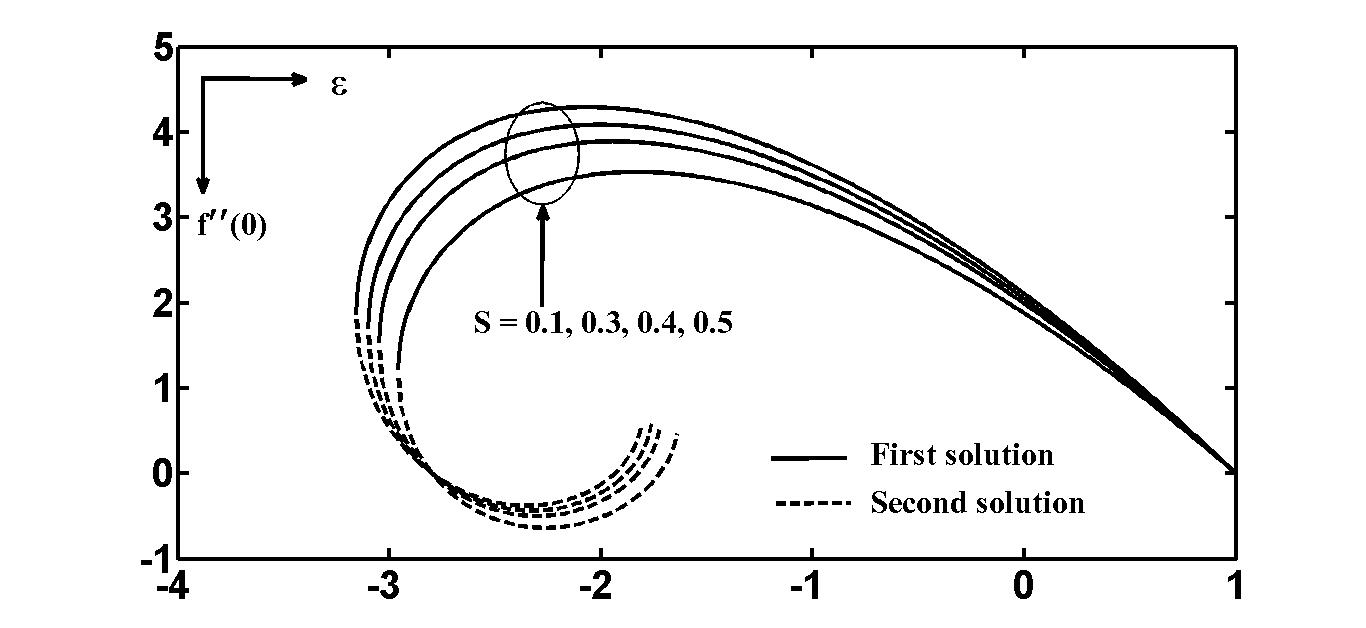}
     \caption{Skin friction for $\lambda=\gamma=0$}\label{1.3.6a}
   \end{subfigure}\\
   \begin{subfigure}{1.2\linewidth} \centering
     \includegraphics[scale=0.3]{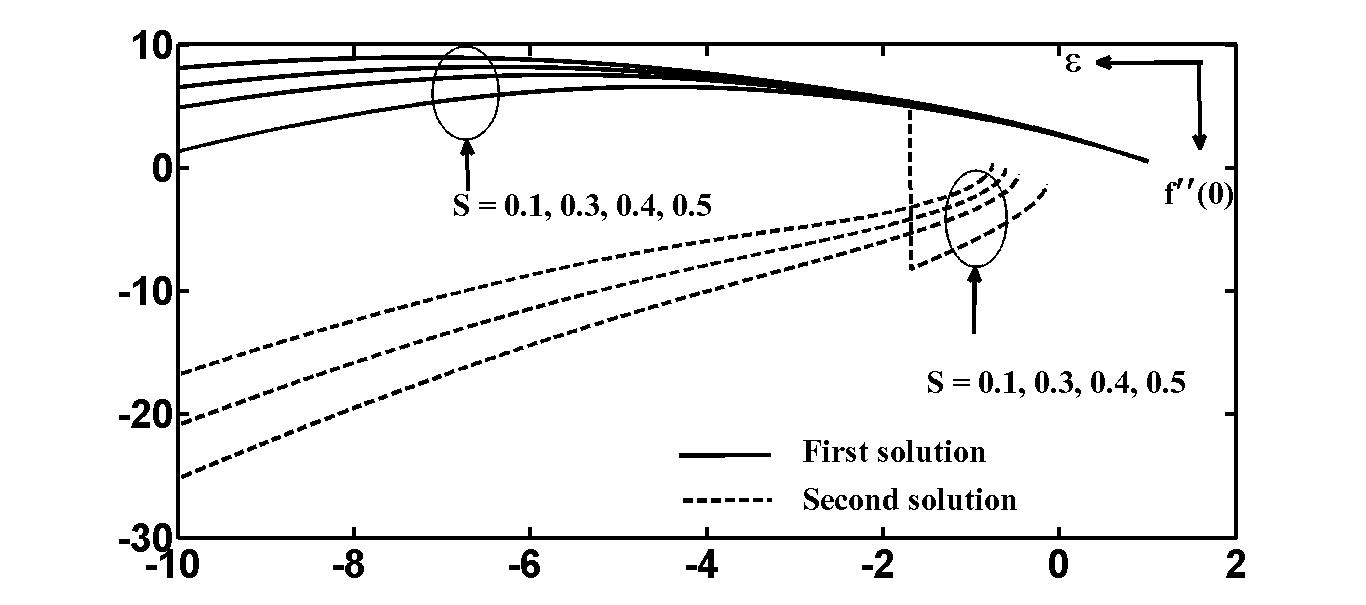}
     \caption{ Skin friction for $\lambda=2,\gamma=0.5$}\label{1.3.6b}
   \end{subfigure}
\caption{Variation of Skin friction with $\epsilon$ for various values of $S$ with $ M=1, K=1, Pr=3$}  \label{fig:1.3.6}
\end{figure}

\begin{figure}[!htbp]
\centering
   \begin{subfigure}{1.2\linewidth} \centering
     \includegraphics[scale=0.3]{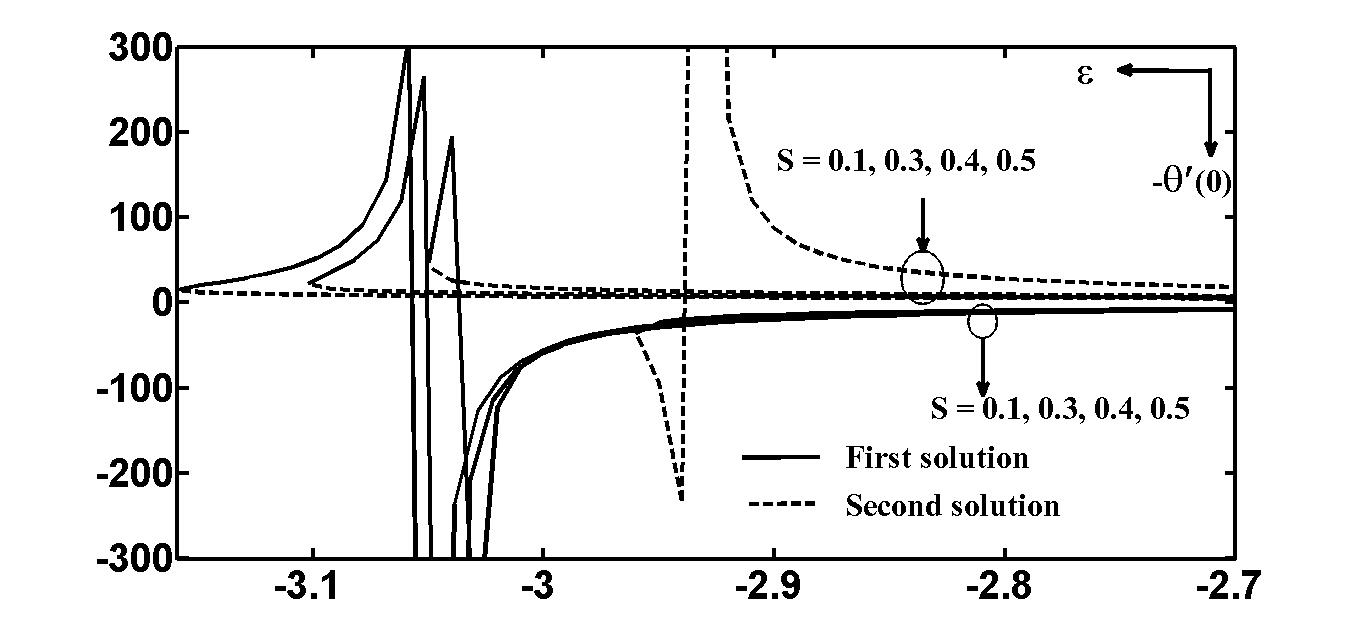}
     \caption{Nusselt number for $\lambda=\gamma=0$} \label{1.3.7a}
   \end{subfigure}\\
   \begin{subfigure}{1.2\linewidth} \centering
     \includegraphics[scale=0.3]{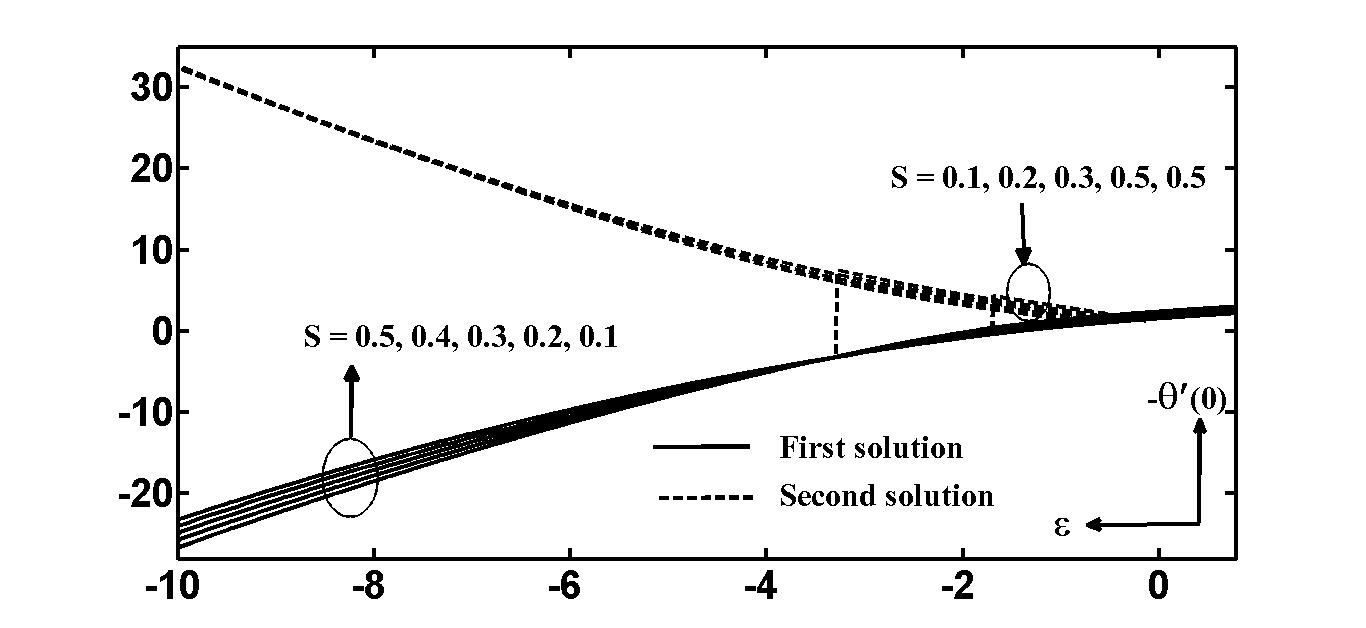}
     \caption{Nusselt number for $\lambda=2,\gamma=0.5$}\label{1.3.7b}
   \end{subfigure}
\caption{Variation of $-\theta'(0)$ with $\epsilon$ for various values of $S$ with $ K=1, M=1,Pr=3$}  \label{fig:1.3.7}
\end{figure}
Moreover, all-inclusive nature of bifurcation points and dual solutions and comparative details can be perceived through Figs. \ref{fig:1.3.2} to \ref{fig:1.3.7}. From these Figures we see that first solution profiles of skin friction $f''(0)$ and Nusselt number $-\theta'(0)$ are enhanced with $M$, $K$ and $S$ for both cases (a) and (b), but for second solutions $f''(0)$ and $-\theta'(0)$ are reduced. The interesting engineering results are obtained when we take $\lambda\ne0$ and $\gamma\ne0$. Case (b) of the mentioned Figs. demonstrates that $f''(0)$ and $-\theta'(0)$ can be computed even for larger shrinking rates and the unique solution range is found to be significantly increased. \\

\par
\begin{figure}[!htbp]
\centering
   \begin{subfigure}{1.2\linewidth} \centering
     \includegraphics[scale=0.3]{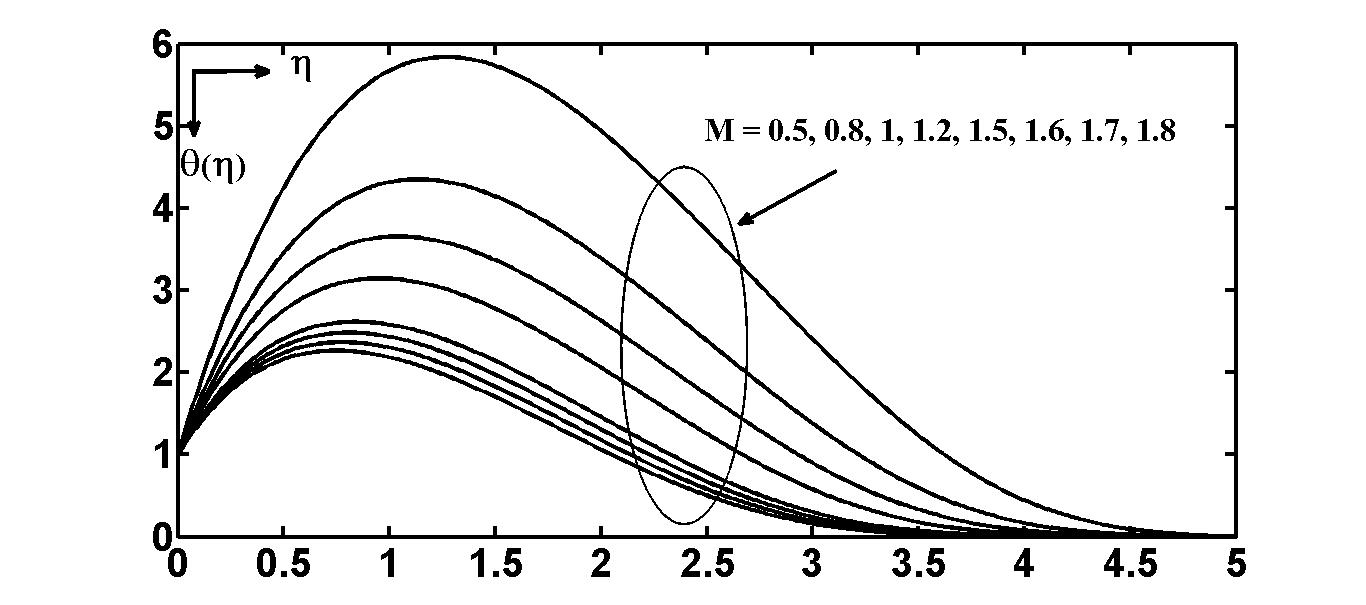}
     \caption{First solution} \label{1.3.9a}
   \end{subfigure}\\
   \begin{subfigure}{1.2\linewidth} \centering
     \includegraphics[scale=0.3]{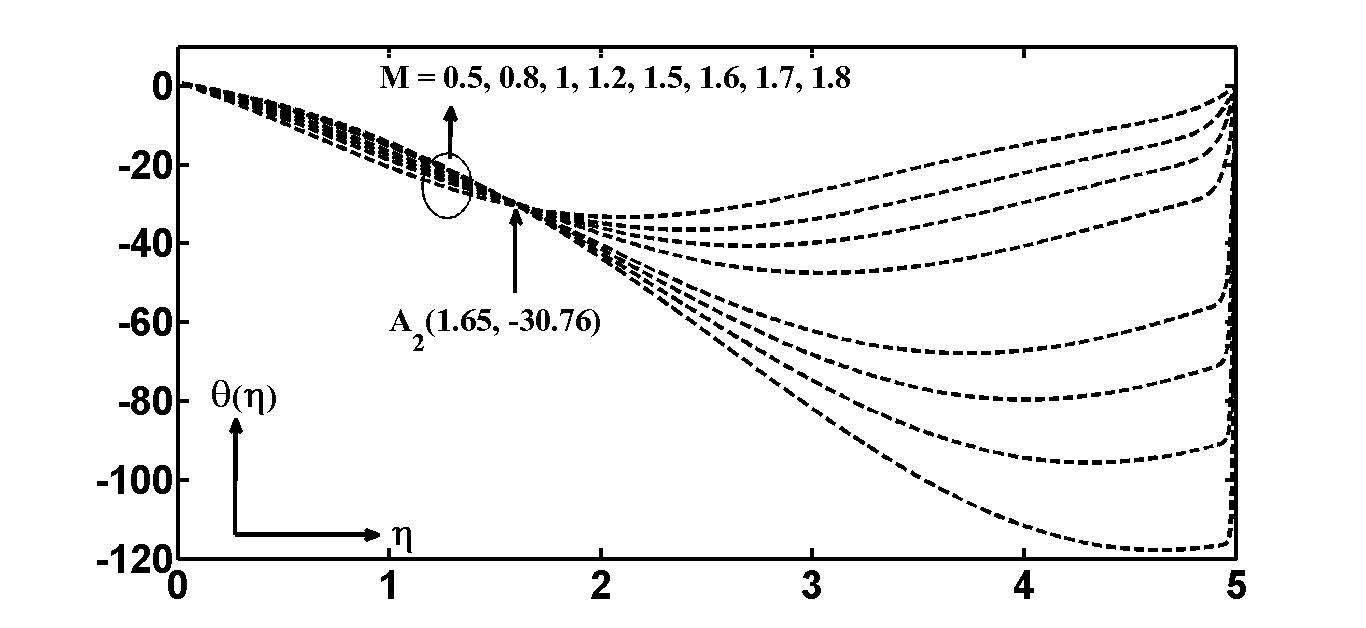}
     \caption{Second solution}\label{1.3.9b}
   \end{subfigure}
\caption{Effect of magnetic parameter ($M$) on Temperature profile}  \label{fig:1.3.9}
\end{figure}

 The effect of the strength of magnetic field $(M)$ on the first and second solution profiles of velocity $f'(\eta)$ and temperature
\begin{figure}[!htbp]
\centering
   \begin{subfigure}{1.2\linewidth} \centering
     \includegraphics[scale=0.3]{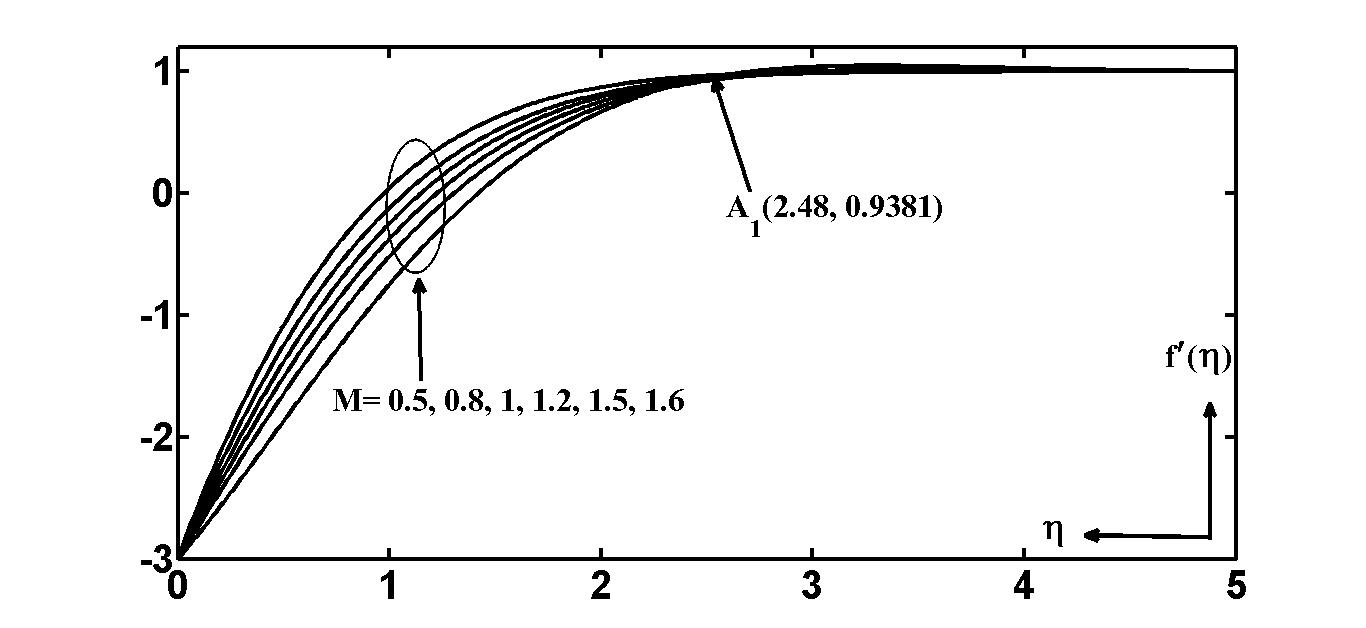}
     \caption{First Solution}\label{1.3.8a}
   \end{subfigure}\\
   \begin{subfigure}{1.2\linewidth} \centering
     \includegraphics[scale=0.3]{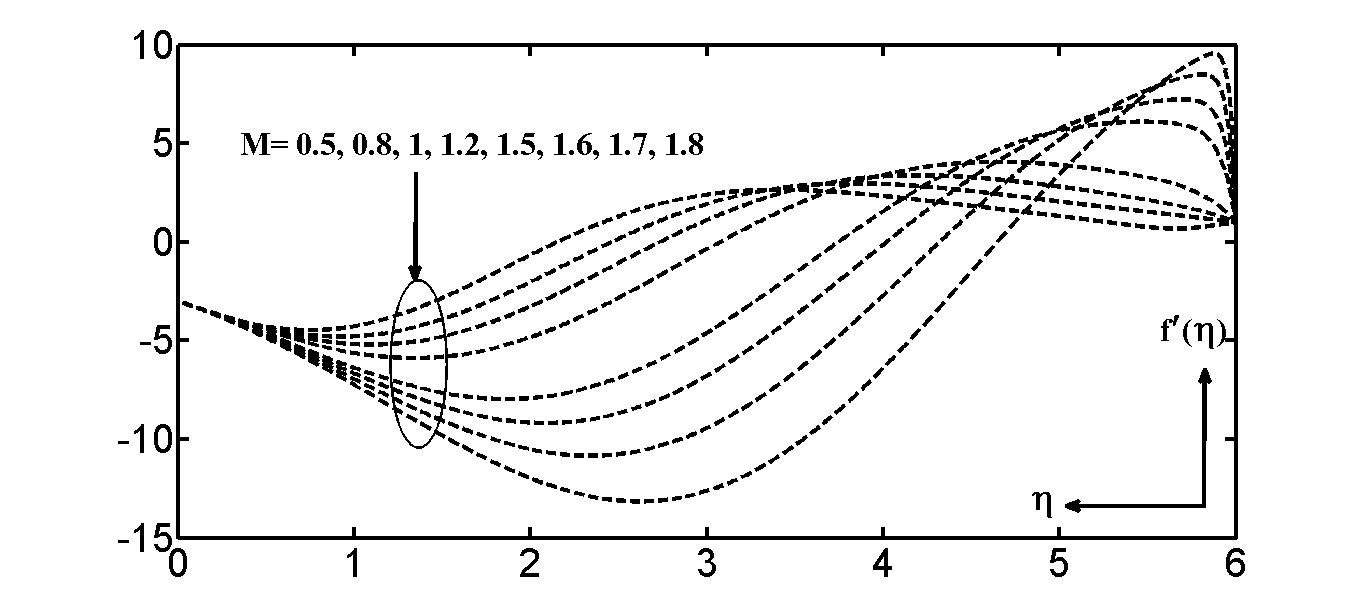}
     \caption{Second solution}\label{1.3.8b}
   \end{subfigure}
\caption{Effect of magnetic parameter ($M$) on Velocity profile}  \label{fig:1.3.8}
\end{figure}
 $\theta(\eta)$ distributions have been presented through Figs. \ref{1.3.8a}, \ref{1.3.8b} and \ref{1.3.9a}, \ref{1.3.9b} respectively. It is inferred from Figs. \ref{1.3.8a} and \ref{1.3.8b} that first solution profiles of $f'(\eta)$ are increased and second solution are decreased with the increasing $M$; and the boundary layer thickness decays for first solution whereas it enhances for second solution. An interception point $A_1(1.69, -31.19)$ has been noted for first branch, however the length of interception points is found to be enhancing with $M$ for second branch. Another interesting feature observed from Fig. \ref{1.3.8b} is that large velocity gradients exist at the far field boundary conditions. These gradients prevail due to the presence of nonlinear convection parameter. The magnetic field has opposite effect on the temperature profiles, that is, profiles of first solution branch are reduced and second solution branch are increased with $M$. Here, an interception point $A_2(1.65, -30.76)$ has been observed for second branch of $\theta(\eta)$ whereas no cross flow points has been detected in the profiles of first branch solutions of temperature.\\

\begin{figure}[!htbp]
\centering
   \begin{subfigure}{1.2\linewidth} \centering
     \includegraphics[scale=0.3]{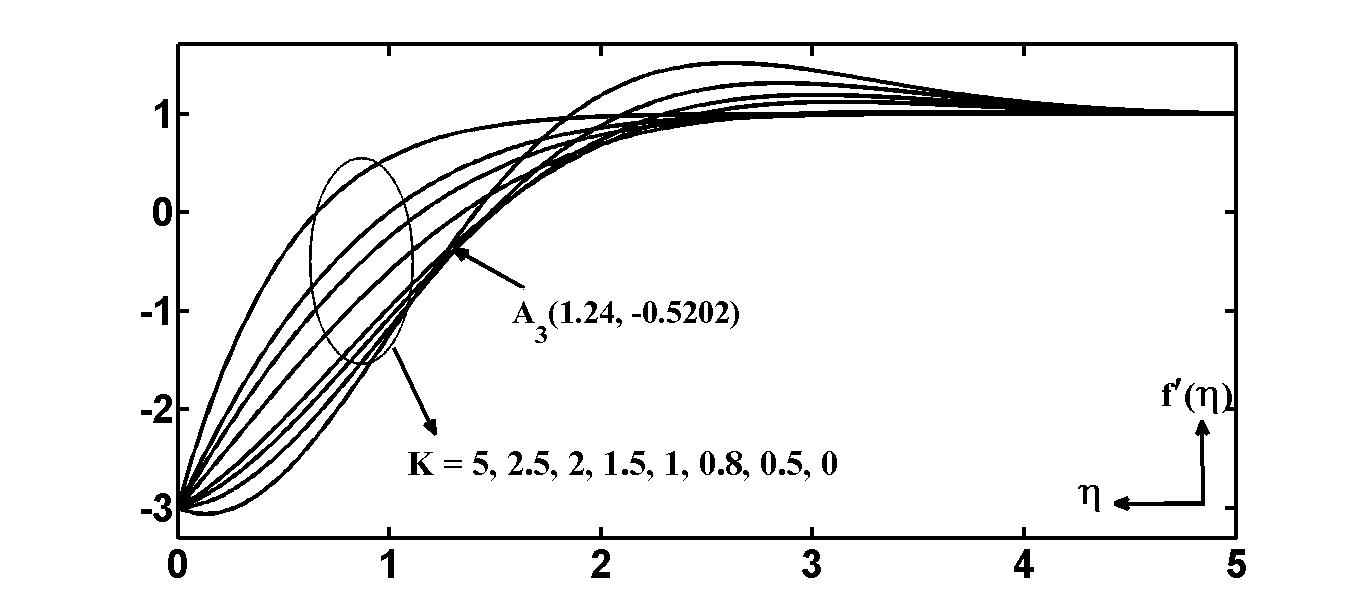}
   \caption{First solution}\label{1.3.10a}
   \end{subfigure}\\
   \begin{subfigure}{1.2\linewidth} \centering
     \includegraphics[scale=0.3]{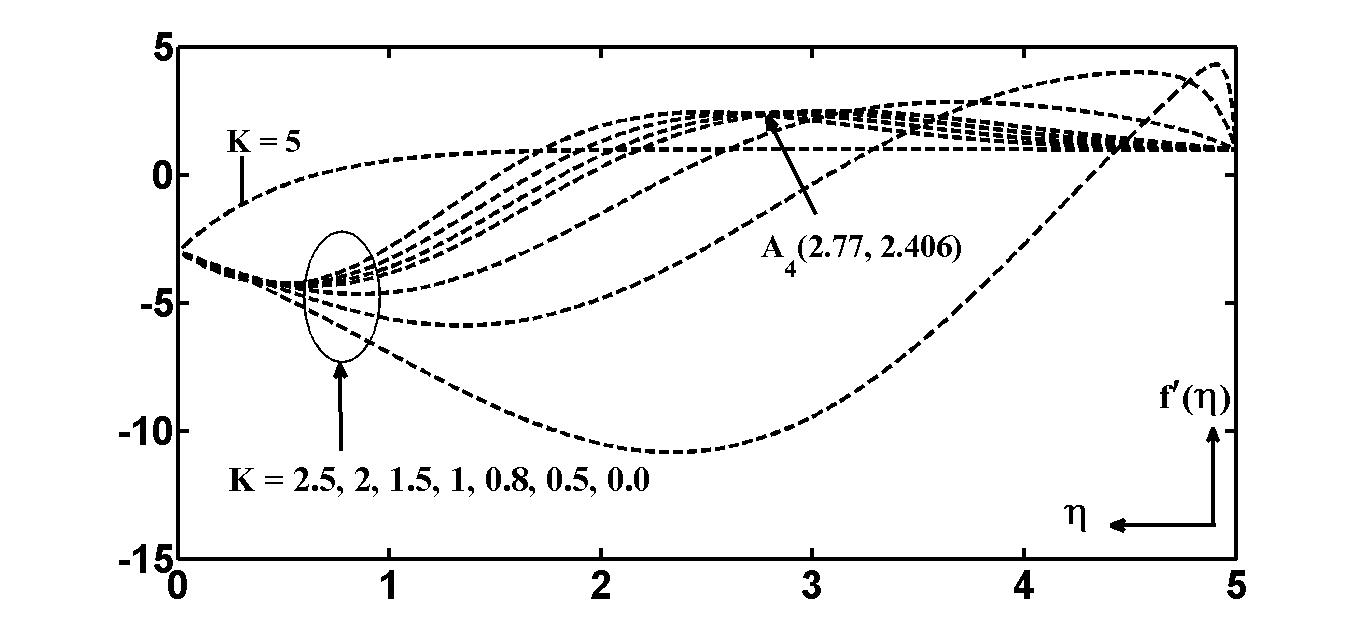}
     \caption{Second  solution}\label{1.3.10b}
   \end{subfigure}
\caption{Effect of Permeability parameter ($K$) on velocity profiles}  \label{fig:1.3.10}
\end{figure}
\begin{figure}[!htbp]
\centering
   \begin{subfigure}{1.2\linewidth} \centering
     \includegraphics[scale=0.3]{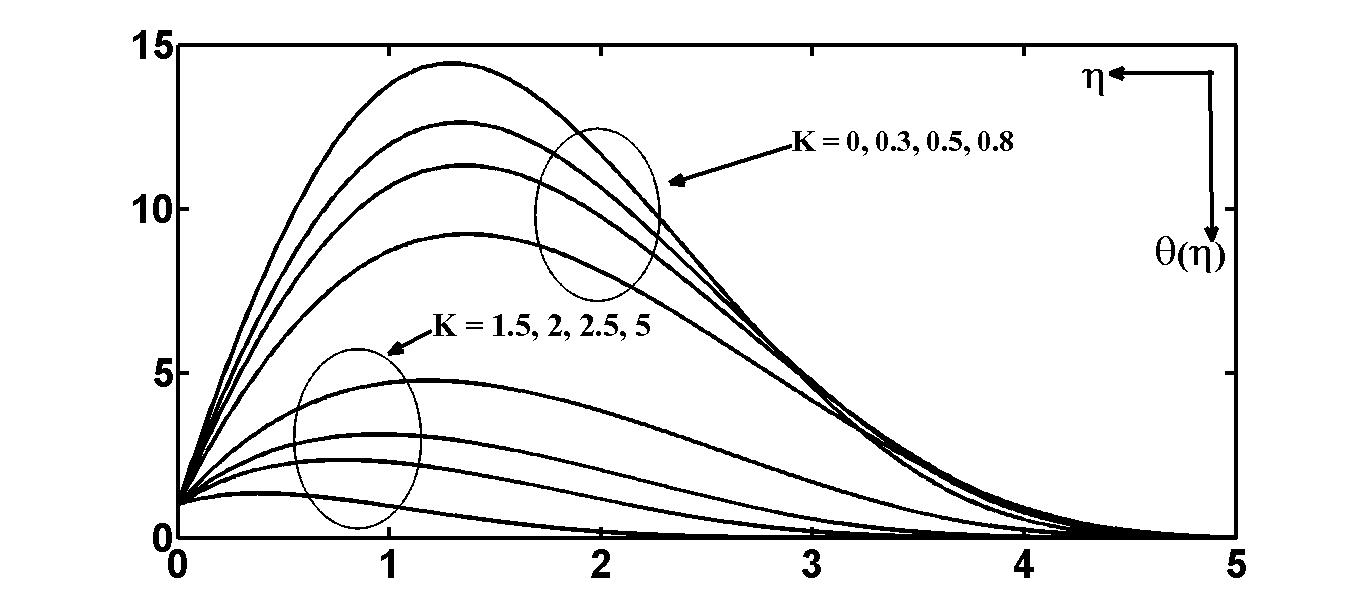}
     \caption{First solution} \label{1.3.11a}
   \end{subfigure}\\
   \begin{subfigure}{1.2\linewidth} \centering
     \includegraphics[scale=0.3]{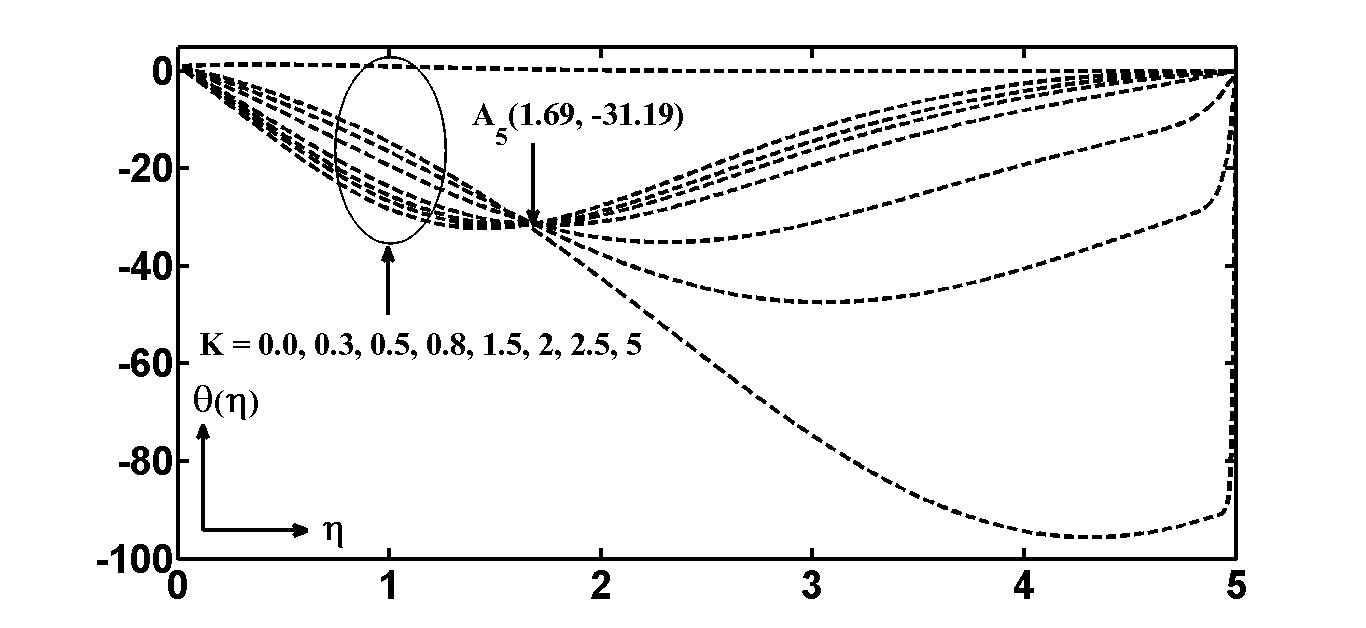}
     \caption{Second solution}\label{1.3.11b}
   \end{subfigure}
\caption{Effect of permeability parameter ($K$) on Temperature profile}  \label{fig:1.3.11}
\end{figure}

Figs. \ref{1.3.10a}, \ref{1.3.10b} and \ref{1.3.11a}, \ref{1.3.11b} have been plotted to demonstrate the influence of porous medium permeability $(K)$ on first and second solution branch profiles of $f'(\eta)$ and temperature $\theta(\eta)$ respectively. The velocity for first increases with $K$ whereas $f'(\eta)$ profiles are decreased for second solution. Two interception points $A_3(1.24, -0.5202)$ and $A_4(2.77, 2.406)$ have been observed for first and second solution branch of $f'(\eta)$ respectively when $K \le1$. However, opposite trends have been illustrated by Figs. \ref{1.3.11a} and \ref{1.3.11b} for temperature profiles with $K$. For the second solution branch of $\theta(\eta)$, a cross flow point $A_5(1.69, -31.19)$ has been noticed; and further large temperature gradients existing at far field boundary condition are reduced with the increasing $K$. The value $K=5$ for porous medium permeability tends to stabilize the velocity and temperature fields.\\

\begin{figure}[!htbp]
\centering
   \begin{subfigure}{1.2\linewidth} \centering
     \includegraphics[scale=0.3]{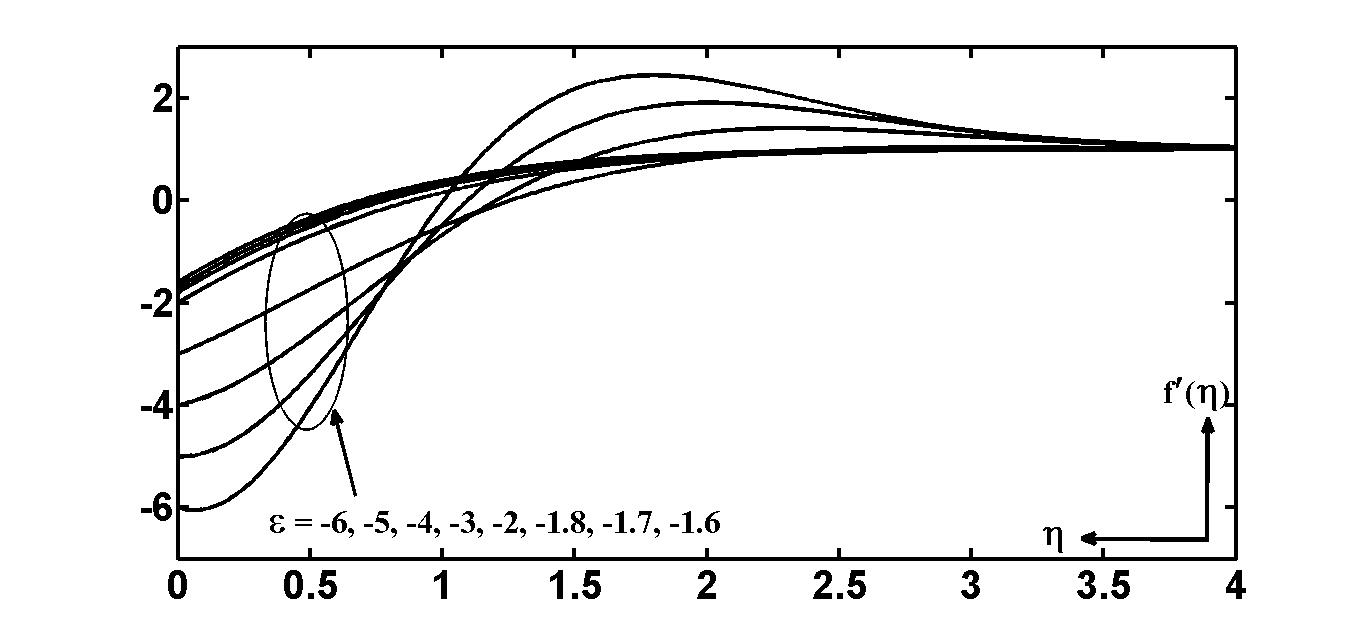}
     \caption{First solution}\label{1.3.12a}
   \end{subfigure}\\
   \begin{subfigure}{1.2\linewidth} \centering
     \includegraphics[scale=0.3]{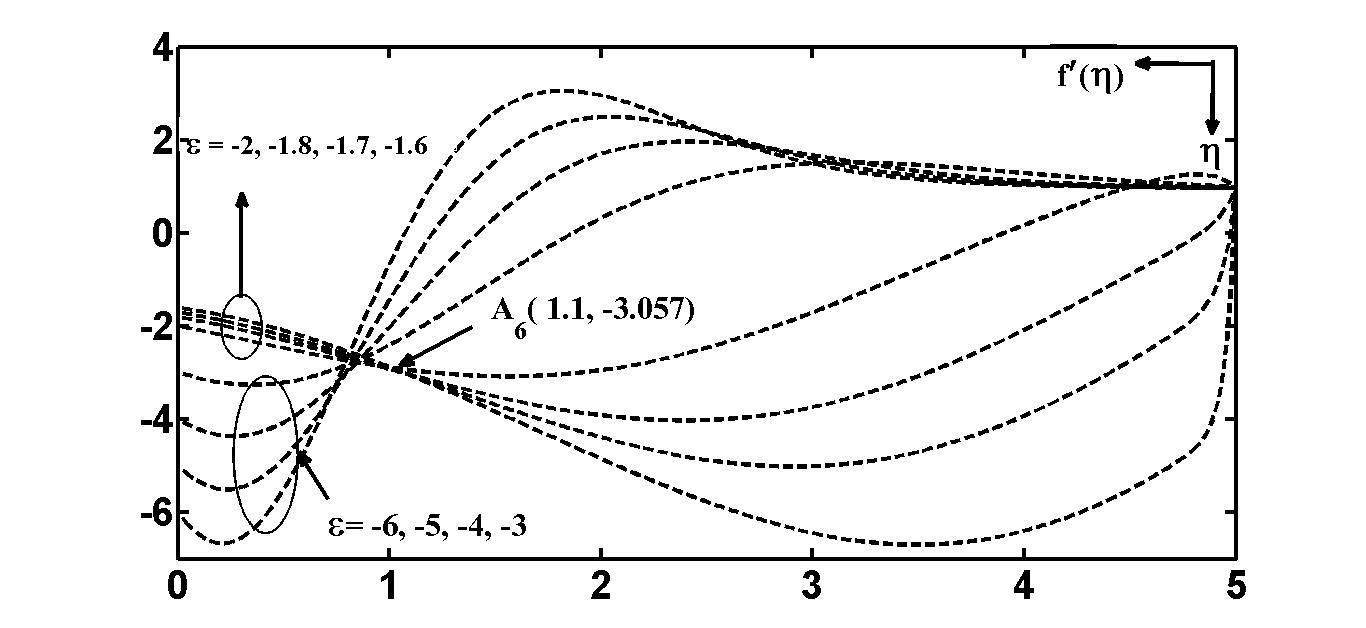}
     \caption{Second solution}\label{1.3.12b}
   \end{subfigure}
\caption{Effect of shrining parameter ($\epsilon$) on velocity profiles}  \label{fig:1.3.12}
\end{figure}

\begin{figure}[!htbp]
\centering
   \begin{subfigure}{1.2\linewidth} \centering
     \includegraphics[scale=0.3]{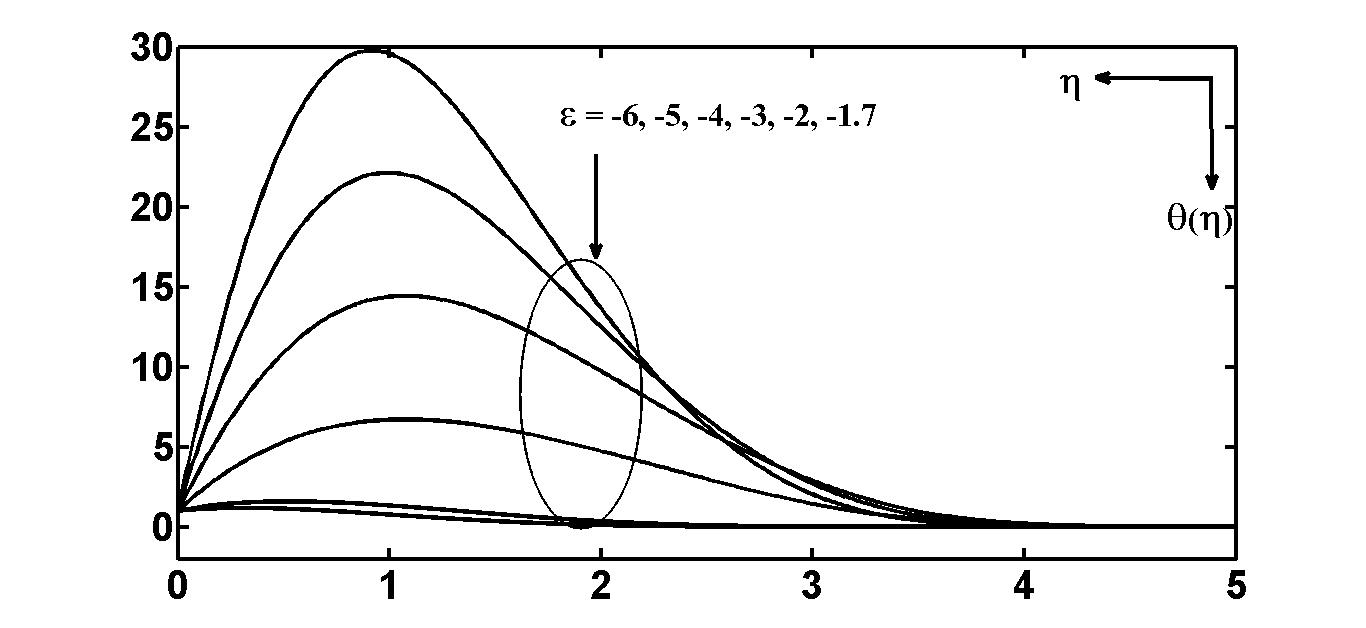}
     \caption{First solution}\label{1.3.13a}
   \end{subfigure}\\
   \begin{subfigure}{1.2\linewidth} \centering
     \includegraphics[scale=0.3]{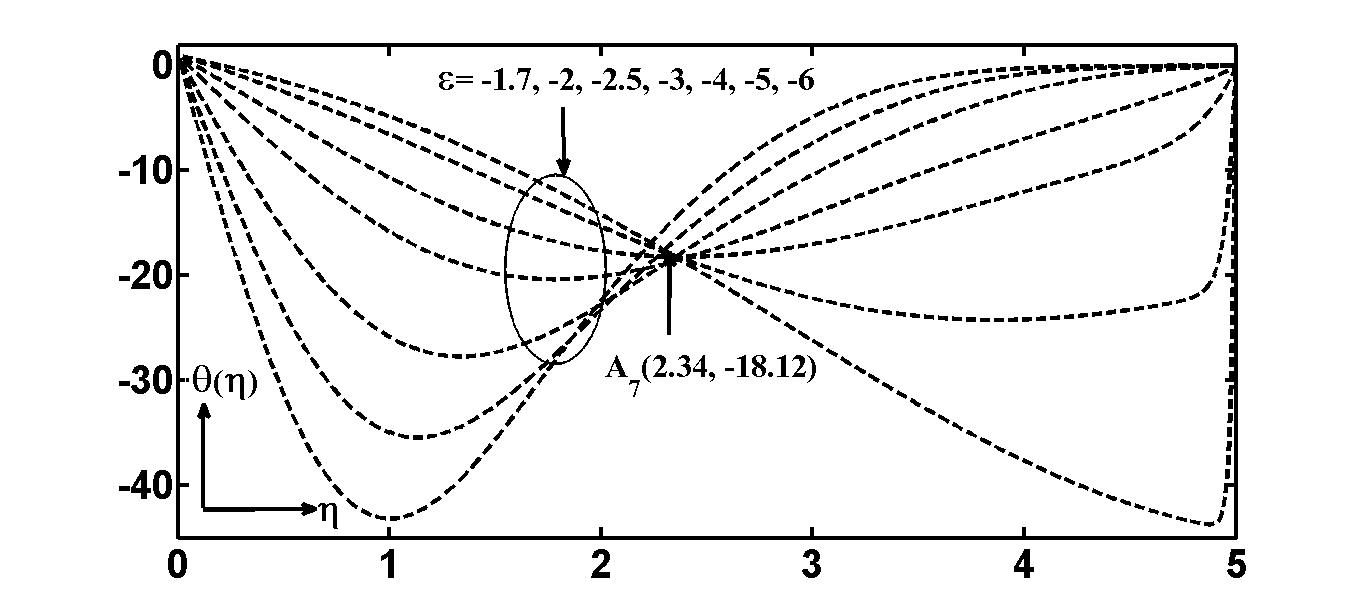}
     \caption{Second solution}\label{1.3.13b}
   \end{subfigure}
\caption{Effect of shrinking parameter ($\epsilon$) on Temperature profile}  \label{fig:1.3.13}
\end{figure}

Figs. \ref{1.3.12a}, \ref{1.3.12b} and \ref{1.3.13a}, \ref{1.3.13b} illustrate the effect of velocity ratio parameter $(\epsilon)$ on the distribution of velocity $f'(\eta)$ and temperature $\theta(\eta)$ respectively. It is
 indicated by Figs. \ref{1.3.12a} and \ref{1.3.12b} that both the solution branches of $f'(\eta)$ are increased with $c/a$ and an interception point $A_6(1.1, -3.057)$ (for $c/a \ge-2$) has been detected for the second solution branch of $f'(\eta)$. Also for second solution, the velocity gradients at far field boundary conditions are again found to be decreasing with $c/a$. The temperature profiles for first solution in Figure \ref{1.3.13a} are reduced with $\epsilon$ and, are enhanced in Fig. \ref{1.3.13b} for second solution. The flow patterns are found to be reversed after crossing the interception point $A_7(2.34, -18.12)$ for $c/a \ge -3$. Here, the profiles of $f'(\eta)$ and $\theta(\eta)$ for second solution also preserve large gradients at far field boundary condition and both are found to be enhancing with the increase of $c/a$.\\
\noindent

\begin{figure}[!htbp]
\centering
   \begin{subfigure}{1.2\linewidth} \centering
     \includegraphics[scale=0.3]{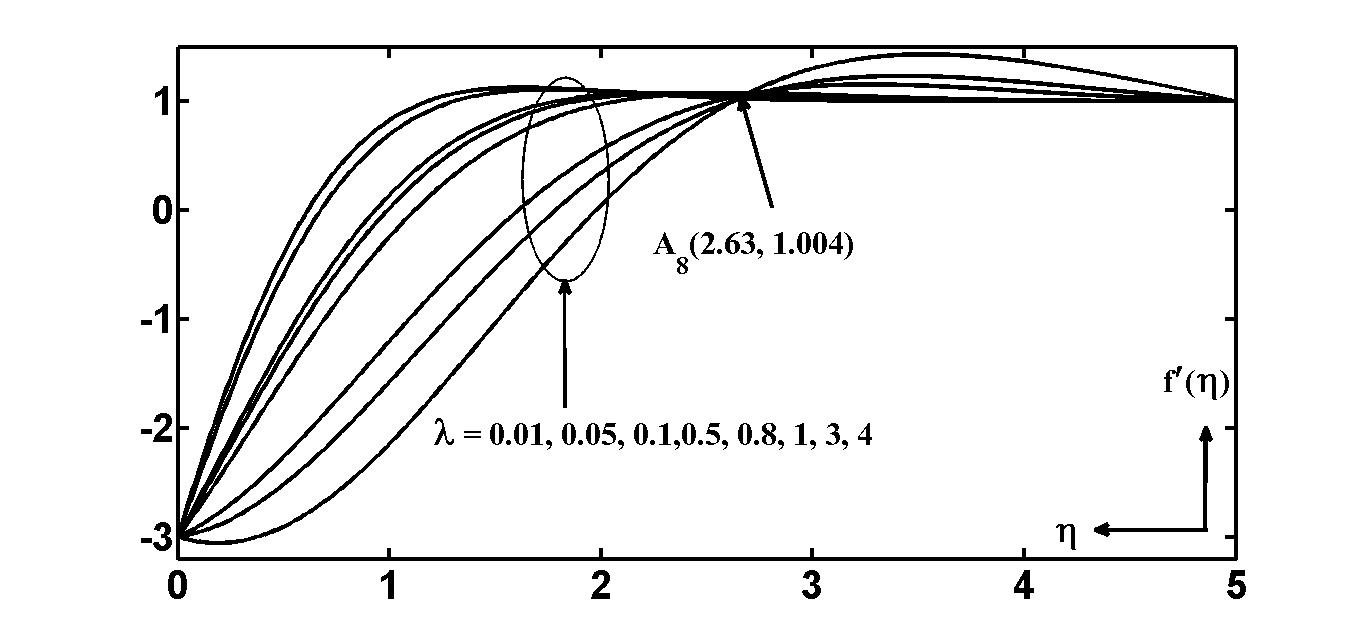}
     \caption{First solution}\label{1.3.14a}
   \end{subfigure}\\
   \begin{subfigure}{1.2\linewidth} \centering
     \includegraphics[scale=0.3]{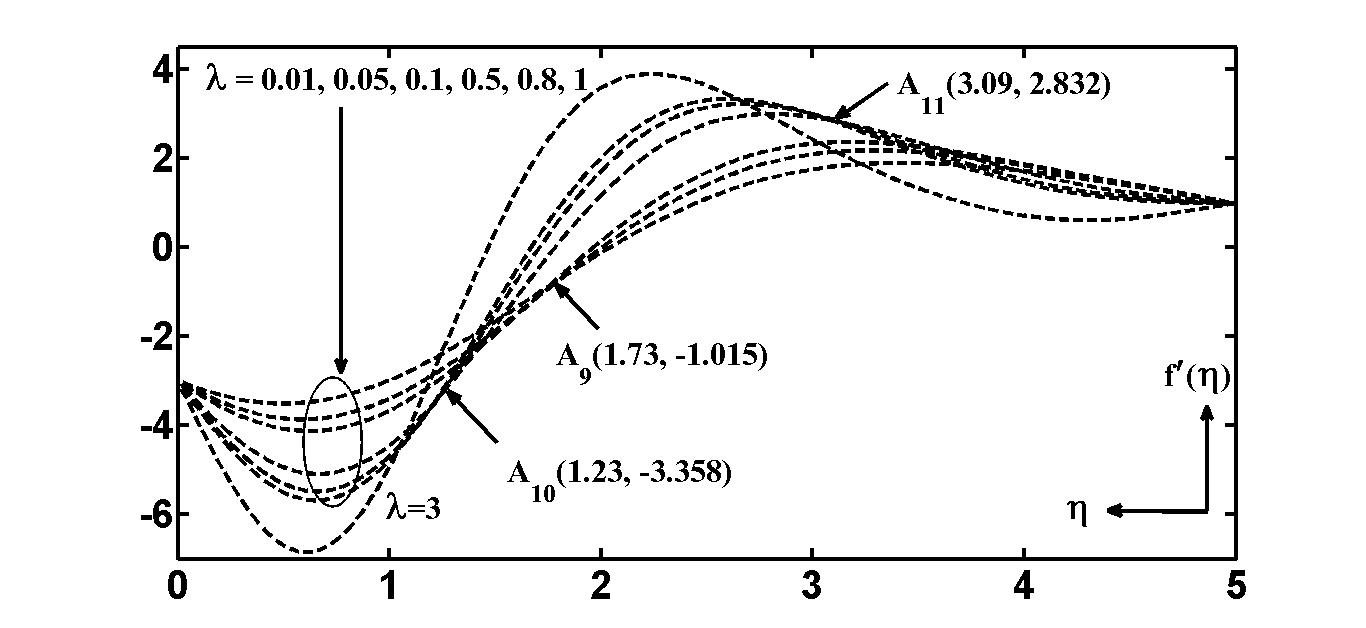}
     \caption{Second solution}\label{1.3.14b}
   \end{subfigure}
\caption{Effect of buoyancy parameter ($\lambda$) on velocity profiles}  \label{fig:1.3.14}
\end{figure}

\begin{figure}[!htbp]
\centering
   \begin{subfigure}{1.2\linewidth} \centering
     \includegraphics[scale=0.3]{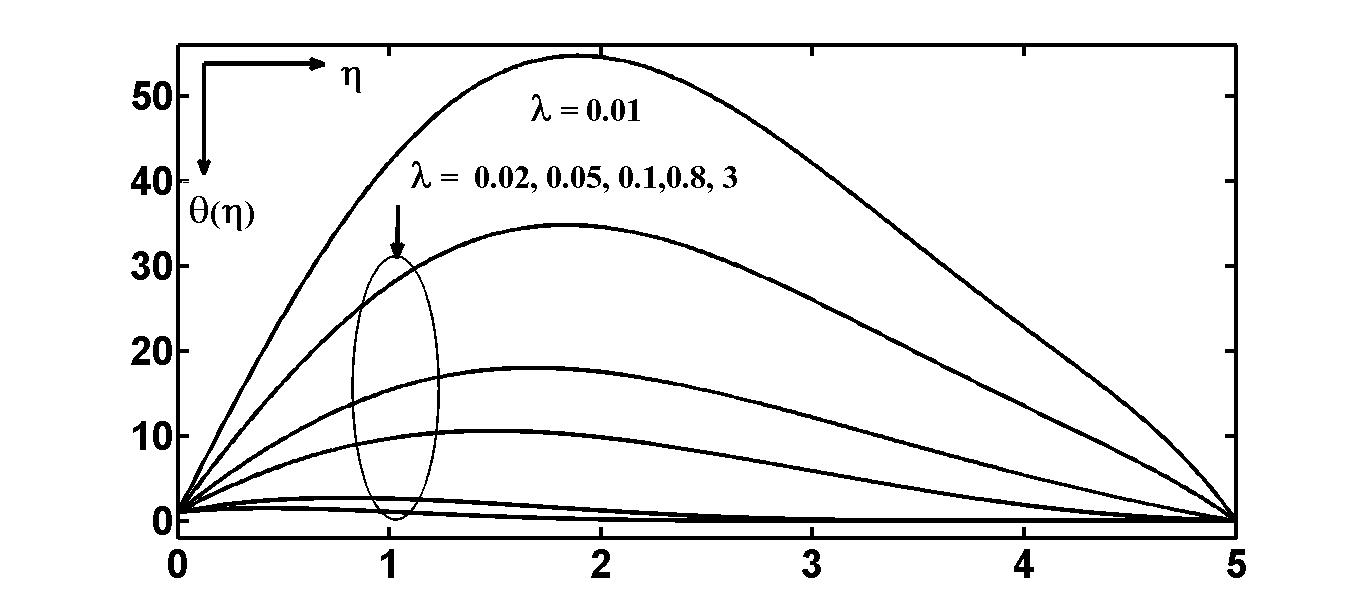}
     \caption{First solution}\label{1.3.15a}
   \end{subfigure}\\
   \begin{subfigure}{1.2\linewidth} \centering
     \includegraphics[scale=0.3]{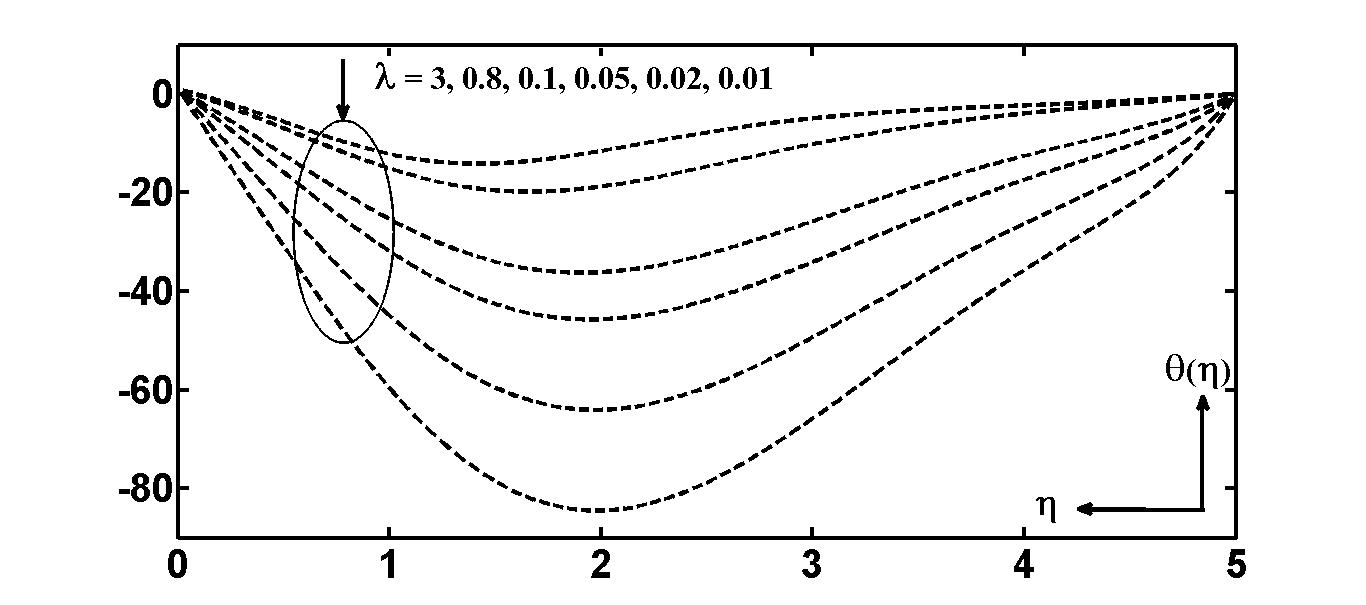}
     \caption{Second solution}\label{1.3.15b}
   \end{subfigure}
\caption{Effect of buoyancy parameter  ($\lambda$) on Temperature profile}  \label{fig:1.3.15}
\end{figure}

In Figs. \ref{1.3.14a}, \ref{1.3.14b} and \ref{1.3.15a}, \ref{1.3.15b}; the effect of buoyancy parameter $(\lambda)$ on the solution branches of velocity $f'(\eta)$ and temperature $f'(\eta)$ has been exhibited respectively. First solution branch of $f'(\eta)$ increases and second branch decreases with the increasing $\lambda$. A critical point $A_8(2.63, 1.004)$ for the first solution branch has been observed at $\eta=2.63$ distance away from the sheet. The most attractive phenomenon is one which is revealed by Fig. \ref{1.3.14b}. It shows that number of interception points are enhanced with the increasing $\lambda$. Further, it is essential to point out here that the second solution branches are representing a backward flow due to the physics of opposite directions of shrinking and straining velocities. However, reverse behavior has been reflected by the first and second solution branch of temperature profiles with respect to $\lambda$. The second solution of $\theta(\eta)$ is always numerically lesser than than the first solution whereas second solution of $f'(\eta)$ crosses its first solution.\\

\begin{figure}[!htbp]
\centering
   \begin{subfigure}{1.2\linewidth} \centering
     \includegraphics[scale=0.3]{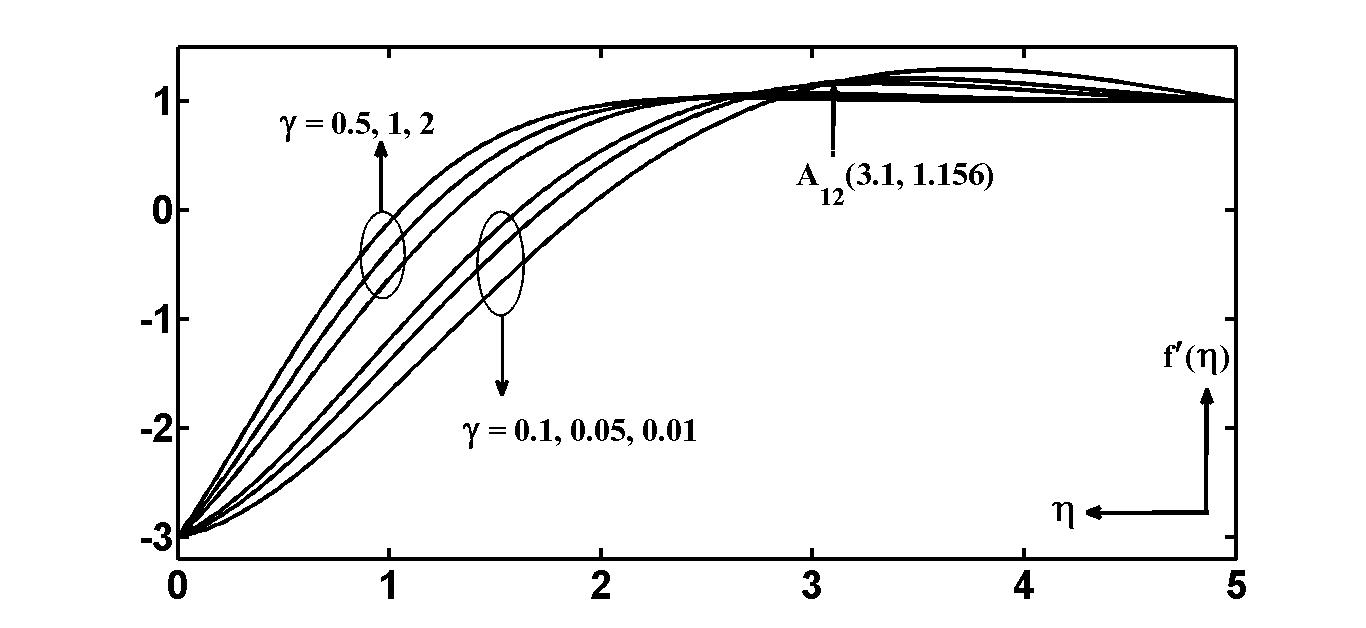}
     \caption{First solution}\label{1.3.16a}
   \end{subfigure}\\
   \begin{subfigure}{1.2\linewidth} \centering
     \includegraphics[scale=0.3]{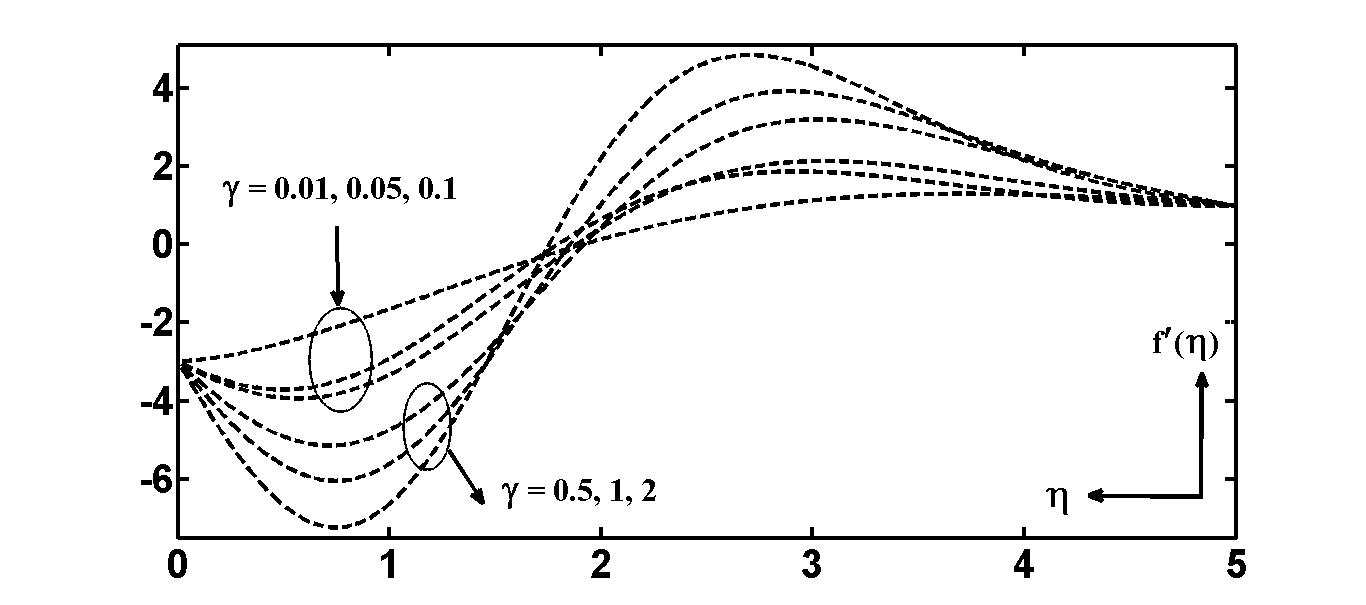}
     \caption{Second solution}\label{1.3.16b}
   \end{subfigure}
\caption{Effect of nonlinear convection ($\gamma$) on velocity profiles}  \label{fig:1.3.16}
\end{figure}

\begin{figure}[!htbp]
\centering
   \begin{subfigure}{1.2\linewidth} \centering
     \includegraphics[scale=0.3]{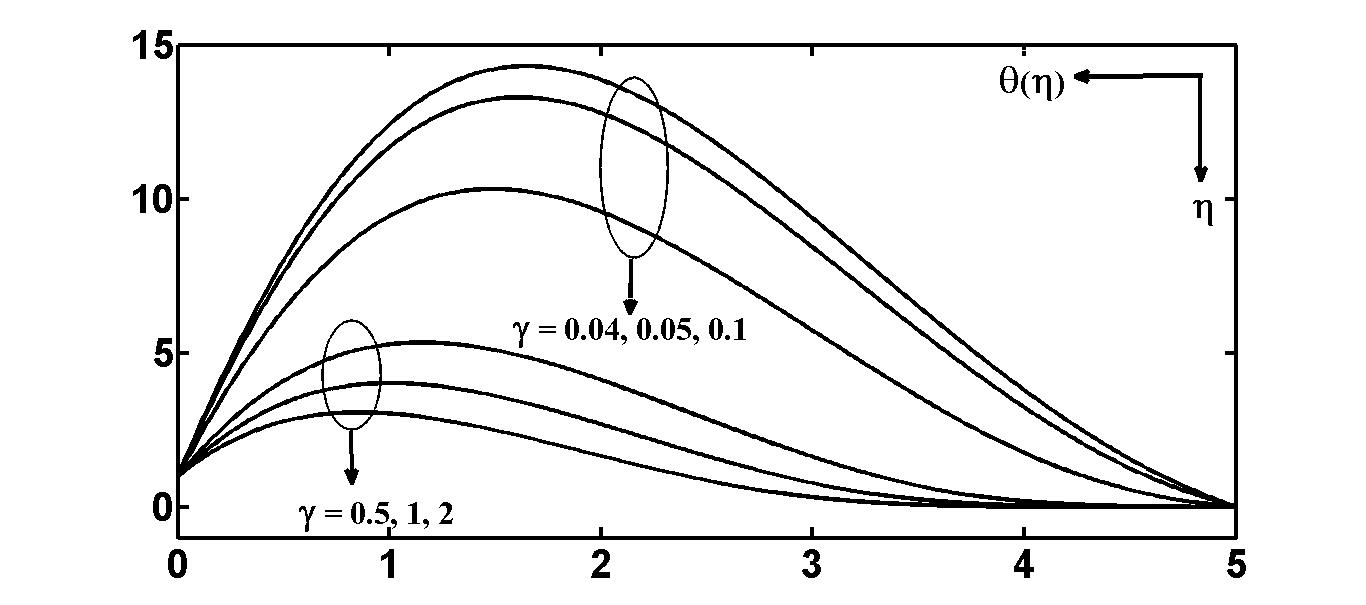}
     \caption{First solution}\label{1.3.17a}
   \end{subfigure}\\
   \begin{subfigure}{1.2\linewidth} \centering
     \includegraphics[scale=0.3]{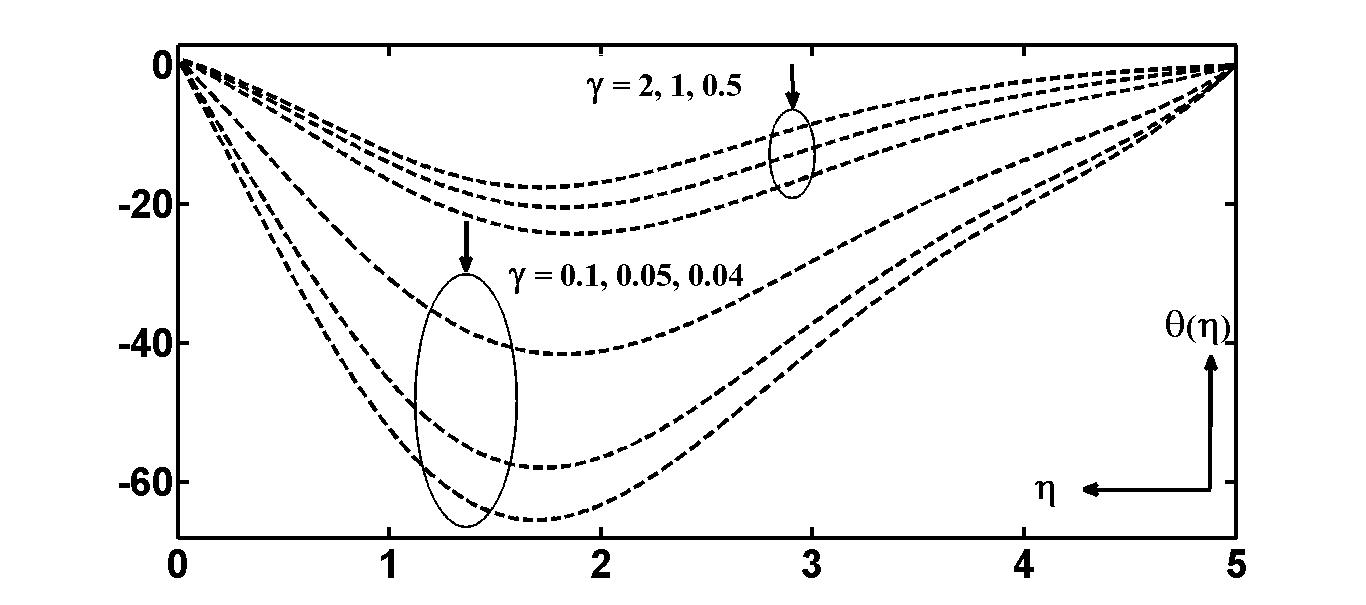}
     \caption{Second solution}\label{1.3.17b}
   \end{subfigure}
\caption{Effect of nonlinear convection ($\gamma$) on Temperature profile}  \label{fig:1.3.17}
\end{figure}

Now, we proceed to discuss the variations of solution profiles of velocity $f'(\eta)$ and temperature $\theta(\eta)$ with respect to nonlinear convection parameter $(\gamma)$ through the Figs\ref{1.3.16a}, \ref{1.3.16b} and \ref{1.3.17a}, \ref{1.3.17b}  respectively. First solution branch of $f'(\eta)$ increases with $\gamma$ whereas second solution of $f'(\eta)$ decreases with it. Moreover, critical points for both the solution branches of $f'(\eta)$ have been observed; and after crossing these points of interest, the behavior is found to be reversed. However, opposite trends have been indicated by the first and second solution branch profiles of $\theta(\eta)$. It is remarkable to mention here that no cross flow points exist for these profiles of $\theta(\eta)$.\\

\begin{figure}[!htbp]
\centering
   \begin{subfigure}{1.2\linewidth} \centering
     \includegraphics[scale=0.3]{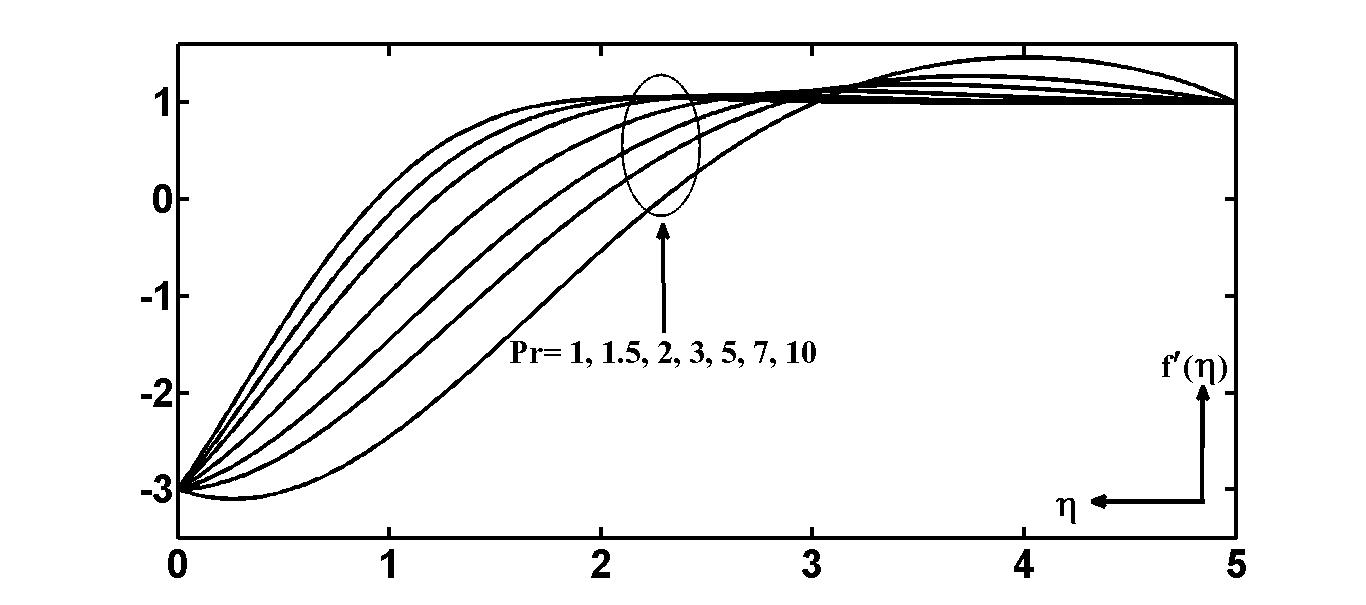}
     \caption{First solution}\label{1.3.18a}
   \end{subfigure}\\
   \begin{subfigure}{1.2\linewidth} \centering
     \includegraphics[scale=0.3]{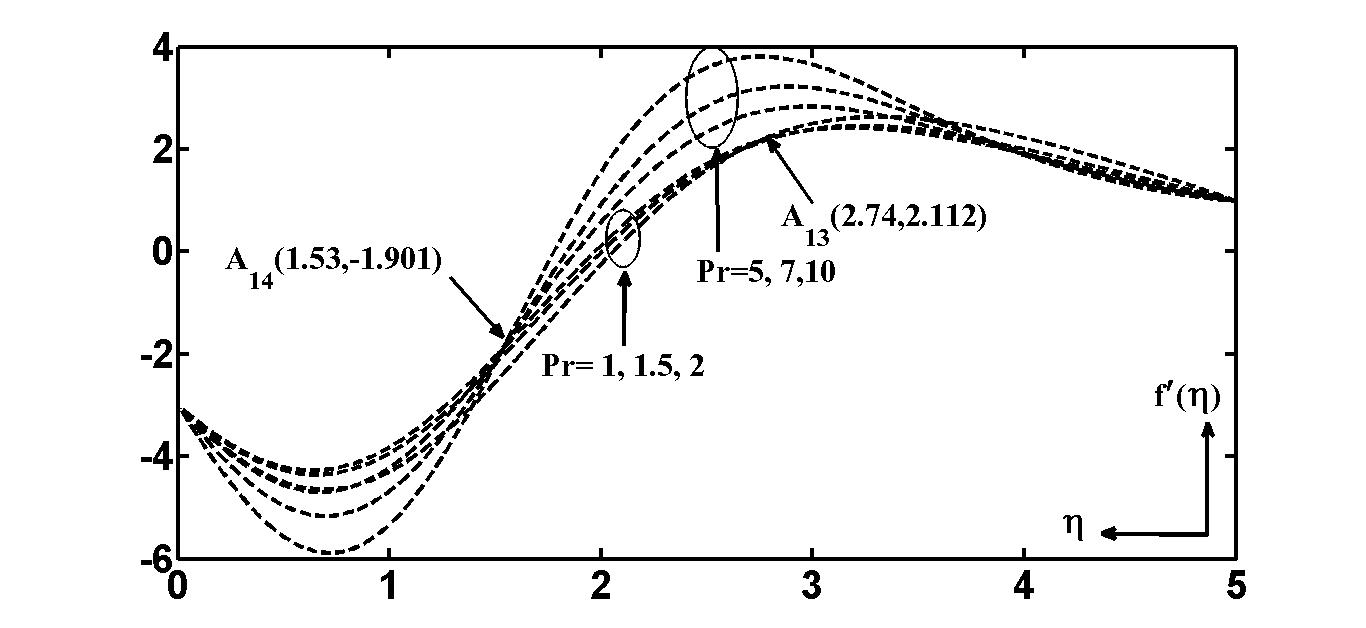}
     \caption{Second solution}\label{1.3.18b}
   \end{subfigure}
\caption{Effect of Prandtl number ($Pr$) on velocity profiles}  \label{fig:1.3.18}
\end{figure}

\begin{figure}[!htbp]
\centering
   \begin{subfigure}{1.2\linewidth} \centering
     \includegraphics[scale=0.3]{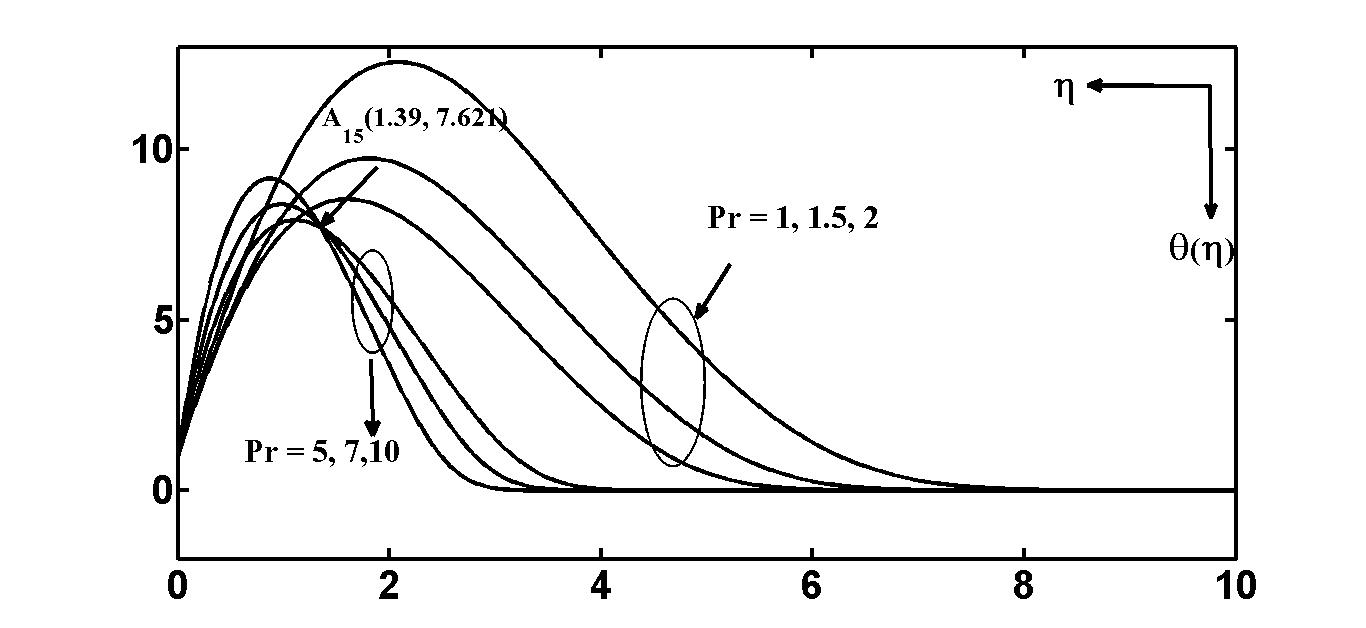}
     \caption{First solution}\label{1.3.19a}
   \end{subfigure}\\
   \begin{subfigure}{1.2\linewidth} \centering
     \includegraphics[scale=0.3]{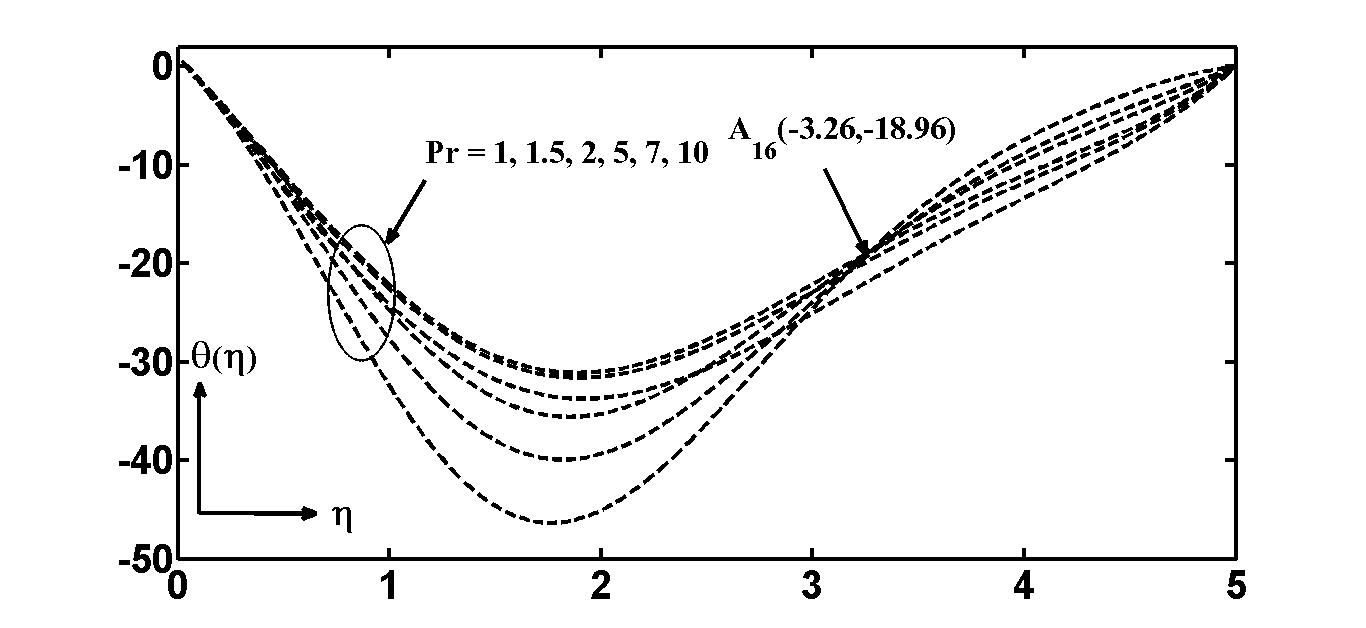}
     \caption{Second solution}\label{1.3.19b}
   \end{subfigure}
\caption{Effect of Prandtl number ($Pr$) on Temperature profiles}  \label{fig:1.3.19}
\end{figure}

The Prandtl number which gives the relative importance of convective heat transfer to conduction heat transfer reveals its impact on the velocity $f'(\eta)$ and temperature $\theta(\eta)$ profiles respectively through the Figs. \ref{1.3.18a}, \ref{1.3.18b} and\ref{1.3.19a}, \ref{1.3.19b} . It is noticed that first solution branch of velocity profiles are increased and the corresponding boundary layer thickness becomes thinner with the increasing $Pr$, however $Pr$ has opposite effect on the second branch solutions, that is, it decreases in the vicinity of the sheet. Multiple interception points have also been detected for second solution branch profiles of $f'(\eta)$. The large values of Prandtl number enhances the velocity profiles as Prandtl number is the ratio of momentum diffusion to thermal diffusion. The first and second branch profiles for $\theta(\eta)$ communicate the same trends as that for $f'(\eta)$ but cross flow points have been observed for $Pr\ge5$ and trends are reversed after crossing these interception points.

\section{Conclusions}
In this investigation, the primary objective was to examine the effect of nonlinear convection $(\gamma)$ on the hydromagnetic stagnation point boundary layer flow and heat transfer due to a vertical and permeable shrinking sheet embedded in porous medium. The numerical dual solutions of the govering self similar equations were obtained by applying implicit finite difference scheme also known as Keller box method. The influence of nonlinear convection parameter and other pertinent parameters on the flow and heat transfer characteristics can be summarized as:
The inclusion of $\gamma$ and interaction among $M$, $K$ and $S$ enhance the solution range significantly, and dual solutions so obtained are found to exist even for larger shrinking rates.
Dual solution range is reduced with the increasing $M$ and $K$ and unique solution range is enhanced significantly if effect of $\gamma$ is considered. Tremendous increase in both the unique and dual solution range has been observed with the increase in $S$.
Nonlinear convection parameter generates steep velocity and temperature gradients at the far field boundary condition, and temperature gradients are reduced with the increasing $M$ and $K$, whereas velocity gradients reveal opposite trends with respect to $M$ and $K$. 
Both velocity and temperature gradients at far field boundary condition for the profiles of second solution are enhanced with the increasing $\epsilon$.
Length of interception points enhances in the second solution profiles of $f'(\eta)$ with the increasing $M$ and $K$, whereas unique interception point has been detected for profiles of $\theta(\eta)$.

\end{document}